\documentclass[prb,preprintnumbers,amsmath,amssymb]{revtex4}

\usepackage[dvips]{graphicx}
\usepackage{amssymb}

\begin{document}

\title{Electron correlations in two-dimensional small quantum dots}

\author{C. Sloggett}
\author{O. P. Sushkov}

\affiliation{School of Physics, University of New South Wales, Sydney
2052, Australia}

\begin{abstract}
We consider circular and elliptic quantum dots with parabolic external
confinement, containing $0 - 22$ electrons and with values of $r_s$ in
the range $0 < r_s < 3$. We perform restricted and unrestricted
Hartree-Fock calculations, and further take into account electron
correlations using second-order perturbation theory. We demonstrate
that in many cases correlations qualitatively change the spin
structure of the ground state from that obtained under Hartree-Fock
and spin-density-functional calculations.  In some cases the
correlation effects destroy Hund's rule.  We also demonstrate that the
correlations destroy static spin-density waves observed in
Hartree-Fock and spin-density-functional calculations.
\end{abstract}

\date{\today}
\maketitle

\section{Introduction}

Recently many theoretical and experimental studies have focused on
quantum dots, in which electrons are confined to a mesoscopically
small region. These structures are of interest both for studying
fundamental physics and for potential nanotechnology applications. 
With advances in materials science, semiconductor quantum dots have
become particularly popular. Most semiconductor dots consist of layers
of semiconductor with the electrons confined by the band structure to
a layer interface, forming a quasi-2D electron gas. 
The electrons are further confined in the plane to form the dot,
usually by charged gates. Quantum dots containing many electrons are
often referred to as ``large dots'', and those with few (often $N <
20$) as ``small dots''.  A large dot has a relatively dense electron
spectrum and therefore poorly controlled external parameters like the
shape of gates, impurities, etc. bring an intrinsic external random
component into any description of the dot.  For a small dot the
uncertainty in the parameters is negligible.  The external confining
potential can be approximated as $U_{\textrm{ext}}=\frac{1}{2} m^*
(\omega_x^2 x^2+\omega_y^2 y^2)$, where $m^*$ is the effective mass of
the electron in the semiconductor, and for circular dots
$\omega_x=\omega_y$.  Such small dots are in essence artificial
atoms. However, they have the advantage of allowing direct measurement
of transport properties, and experimental control over basic
parameters such as the size and strength of the confining potential
and the number of electrons in the dot. They can also be driven by
external magnetic fields that would be inaccessible in atomic physics
with realistic magnetic fields.

Small circular quantum dots have been investigated experimentally by
Tarucha \emph{et al.} \cite{tarucha96} and by Kouwenhoven \emph{et
al.}\cite{kouwenhoven97}.  These studies have clearly demonstrated the
shell structure of the energy spectrum, the effect which one would
expect from analogy with atoms or nuclei.  Where there is a shell
structure we might also expect Hund's rule to be obeyed for the total
spin of the dot.  Quantum dots have attracted the attention of numerous
theoretical studies, for reviews see Refs. \cite{alhassid00,Reimann}.
The electronic structure of 2D small quantum dots has been studied
using spin density functional theory
\cite{KMR,lee98,RKHLM,HW,jiang01,YBB} as well as by restricted and
unrestricted Hartree-Fock methods \cite{YL}.  These studies have also
clearly demonstrated the shell structure of circular dots and they
have confirmed the validity of Hund's rule for such dots. As
expected, elliptical deformation destroys both the shell
structure and Hund's rule.  The spin-density-functional calculations
also lead to spin-density wave states.  However, as has been
pointed out by Hirose and Wingreen \cite{HW} such states are artifacts
of broken spin symmetry in density-functional theory.  This is quite
similar to what has been known to occur in unrestricted Hartree-Fock since
the work of Overhauser \cite{Ov}.

As mentioned, there are similarities between small quantum dots and
atoms. Therefore theoretical methods used in atomic physics could be
efficient in studies of quantum dots. It is known that for few (2-4)
electron atoms the best available method is the pair equation
method. The configuration-interaction method also works well in this
case. However for multi-electron atoms the best available method is
Hartree-Fock with further account taken of correlations using
many-body perturbation theory. This method is used nowadays in
practically all precise calculations for multi-electron atoms, see
e.g. Refs \cite{Dzuba89, Blundell90, Johnson04}, but to the best of
our knowledge it has never been used before for quantum dots. In the
present work we perform such calculations.  Our results for total
energies and shell structure are rather similar to already known ones.
However the results concerning the spin structure are different. We
never observe a spin density wave in a state with total spin zero. The
wave can appear at the Hartree-Fock level, but taking correlations
into account restores the rotational symmetry. So, on this we agree
with works \cite{Ov,HW} and disagree with \cite{KMR,YL}. Studies have
also been done in which the rotational symmetry has been explicitly
restored after unrestricted Hartree-Fock via projection
techniques.\cite{YL2}

In the circular dot we have found that the total spin sometimes varies
from that predicted by Hund's rule. For $N=10$, $15$, $16$, and
$17$, we found that the ground states have minimal spin as opposed to
the previous result of maximal spin within the confines of Hund's
rule.  We have also studied elliptically deformed small dots, paying
special attention to the case $\omega_x/\omega_y=2$. It is well known
from nuclear physics \cite{BM} that the shell structure is restored
when the ratio of frequencies is a rational number.  This is certainly
true for a parabolic potential. For a quantum dot the total
self-consistent potential is not parabolic (even if the external
potential is parabolic), but nevertheless we show that there are some
peculiarities in the spectrum of the dot at $\omega_x/\omega_y=2$.

The structure of the paper is as follows. In Section
\ref{sec:estimates} we present simple semiclassical estimates which
allow us to relate the number of electrons, the strength of the
confining potential, and the relative strength of the Coulomb
interaction.  Section \ref{sec:method} is devoted to an overview of
the computational method.  Section \ref{sec:twobody} concerns the two-
and six-electron problems in the external parabolic potential.  The
two-electron problem can easily be solved exactly, while for the
six-electron problem we compare to the configuration-interaction study
of Reimann \emph{et al.}\cite{Reimann00}. Comparing these accurate
energies with our restricted and unrestricted Hartree-Fock results we
demonstrate how unrestricted Hartree-Fock generates the spin density
waves and how correlations restore the rotational symmetry.  Based on
the results of this section it appears that the method gives reliable
results to at least $r_s \leq 3$, although there appears to be some
N-dependence in the accuracy. Finally, Section \ref{sec:round}
presents our results for circular parabolic dots, and Section
\ref{sec:elliptical} presents results for elliptical dots.

\section{Analytical estimates of parameters}
\label{sec:estimates}

In this section we use atomic units for energy and length
\begin{eqnarray}
\label{units}
E_h^* &=& \frac{m^* e^4}{(4 \pi \epsilon \hbar)^2}
\approx 11 \textrm{meV} \ ,\nonumber\\
a^*_B&=& \frac{4 \pi \epsilon \hbar^2}{m^* e^2}\approx
102 \textrm{\AA} \approx 10nm \ .
\end{eqnarray}
Here $e$ is the elementary charge, $m^*$ is the electron effective
mass, and $\epsilon=\kappa\epsilon_0$ is the dielectric constant of the
material.  Numerical values are presented for $m^* = 0.067 m_e$ and
$\epsilon = 12.9 \epsilon_0$, which correspond to the GaAs commonly used
in quantum dots.  The confining external potential is of the form
\begin{equation}
\label{conf}
U_{ext}=\frac{1}{2}(\omega_x^2 x^2+\omega_y^2 y^2) \to
\frac{1}{2}\omega^2 r^2 \ .
\end{equation}
Assuming that the number of electrons in the dot is large, $N \gg 1$,
we can use the standard semiclassical expression for number density in a 2D
circular dot of radius $R$ \cite{Sneddon}
\begin{equation}
\label{dens}
n(r)=n_0\sqrt{1-r^2/R^2} \ ,
\end{equation}
where $n_0=3N/(2\pi R^2)$.  The electrostatic potential energy at the
origin and at the edge of the dot is
\begin{eqnarray}
\label{UU}
U(0)&=&\frac{\pi^2}{2}n_0 R \ , \nonumber\\
U(R)&=&\frac{\pi^2}{4}n_0 R \ .
\end{eqnarray}
In the Thomas-Fermi approximation the number density at $r=0$ is related
to the Fermi momentum at this point
\begin{equation}
\label{tf1}
p_F^2=2\pi n_0 \ ,
\end{equation}
and the self-consistency equation reads
\begin{equation}
\label{tf2}
\epsilon_F=\frac{p_F^2}{2}=\left.\left(U_{ext}+U\right)\right|_{r=R}-
\left.\left(U_{ext}+U\right)\right|_{r=0} \ .
\end{equation}
At $N \gg 1$ the Fermi energy is small ($\propto N^{- 1/3}$) compared
to the self-consistent potential. So Eq. (\ref{tf2})
implies that $\left.\left(U_{ext}+U\right)\right|_{r=R}=
\left.\left(U_{ext}+U\right)\right|_{r=0}$.  This gives the standard
semiclassical relation between the size of the dot and the number of
electrons (see for example Ref. \cite{KS})
\begin{eqnarray}
\label{rel1}
R&=&\left(\frac{3\pi N}{4\omega^2}\right)^{\frac{1}{3}} \ , \nonumber\\
n_0&=&\left(\frac{6}{\pi^5}\right)^{\frac{1}{3}}\omega^{\frac{4}{3}}N^{\frac{1}{3}} \ .
\end{eqnarray}
The number density of electrons varies from point to point. Therefore
the interaction is less important at the centre and more
important towards the edge of the dot.  To characterize the interaction we
define an average $r_s$ according to the equation
\begin{equation}
\label{rs}
\pi r_s^2 N= \pi R^2 \ ,  \ \ \ \ \  \textrm{or} \ \ \ \ \ R=r_s\sqrt{N} \ .
\end{equation}
Then, using (\ref{rel1}) one finds
\begin{equation}
\label{r1}
r_s= \left(\frac{3\pi}{4}\right)^{\frac{1}{3}} \omega^{-\frac{2}{3}}N^{-\frac{1}{6}}\ .
\end{equation}
Note that this definition of $r_s$ differs from that accepted in
Ref. \cite{Reimann} by a factor of $\left(3\pi/4\right)^{\frac{1}{3}}
\approx 1.33$.  This difference is due to the semiclassical estimate
we use for the dot density in Eq. (\ref{dens}), as opposed to the
assumption\cite{Reimann} that the density is roughly flat.

The total energy of the dot scales as
\begin{equation}
\label{eto}
E \propto N^{\frac{3}{2}} r_s^{-1} \ .
\end{equation}
We will use Eqs. (\ref{rel1}), (\ref{rs}), (\ref{r1}), and (\ref{eto})
not only for circular dots, but for elliptic dots as well. In this
case we will take $\omega =\sqrt{\omega_x \omega_y}$, $R=\sqrt{ab}$.

We stress once more that Eqs. (\ref{rel1}), (\ref{rs}), (\ref{r1}),
and (\ref{eto}) are written in dimensionless atomic units.  Using
Eq. (\ref{units}) one can easily convert them to other units. For
example a GaAs quantum dot with $r_s = 1$ and containing about 20
electrons should have a radius of the order of $R \approx 500
\textrm{\AA} \approx 0.1 \mu \textrm{m}$. The confining potential
$\hbar \omega$ must be on the order of 8 meV once any screening
effects from the surrounding layers are considered.

\section{Method}
\label{sec:method}

\subsection{The model}

The electron wavefunctions are modeled on a square lattice. 
Essentially we consider an Anderson model with Hamiltonian
\begin{widetext}
\begin{equation}
\label{H}
H=\sum_{i\sigma} \left( 4+\frac{1}{4}(\overline{\omega}_x^2 x_i^2+
\overline{\omega}_y^2 y_i^2)\right)
c_{i\sigma}^{\dag}c_{i\sigma}
-\sum_{\langle ij\rangle\sigma} c_{i\sigma}^{\dag}c_{j\sigma}
+\frac{1}{2} \sum_{ij\alpha\beta} \frac{q^2}{|{\bf r}_i-{\bf r}_j|}
c_{i\alpha}^{\dag}c_{i\alpha}
c_{i\beta}^{\dag}c_{i\beta} \ .
\end{equation}
\end{widetext}
Here $c_{i \sigma}^{\dag}$ is the creation operator of an electron
with spin projection $\sigma$ ($\uparrow$ or $\downarrow$) at site $i$
of the two-dimensional square lattice with lattice spacing equal to
one.  $\langle ij \rangle$ represents a sum over nearest neighbor
sites.  The single particle dispersion corresponding to the
Hamiltonian (\ref{H}) is
\begin{equation}
\label{disp}
\epsilon_p=4-2\cos p_x -2\cos p_y \approx p^2 \ .
\end{equation}
At larger momenta the dispersion deviates from quadratic dispersion,
so to simulate the real quadratic dispersion we need to have many
lattice points within one period of the electron wave function, which
is on the order of one atomic unit.  For this reason, the units of
length used in the Anderson Hamiltonian (\ref{H}) are not atomic
units. The Hamiltonian is written in units in which the lattice
spacing is one, where the lattice spacing needs to be significantly
smaller than one atomic unit.

The Coulomb interaction $q^2/|{\bf r}_i-{\bf r}_j|$ is singular at
$i=j$.  This is an unphysical singularity as, firstly, if the integral
were performed in continuous instead of discrete 2D space there would
be no divergence, and secondly as in a real dot there is transverse
(orthogonal to the plane) confinement with some finite effective
length, typically on the order of $10 nm$.

Here, we avoid adding a finite width to our model in order to be able
to compare our results to exact and configuration-interaction 2D
calculations. Instead we consider that when ${\bf r}_i = {\bf r}_j$,
the effective distance between electrons should not be taken to be
zero, but some finite value characteristic of the size of the lattice
cell. We obtain an approximate value for the integral at ${\bf r}_i =
{\bf r}_j$ by treating the wavefunctions as constant over the unit
cell and taking the integral of $q^2/|{\bf r}_i-{\bf r}_j|$ over the
cell. That is, when ${\bf r}_i = {\bf r}_j$, we take

\begin{equation}
\frac{1}{|{\bf r}_i-{\bf r}_j|} = \int_\textrm{lattice cell} \frac{1}{|{\bf r}-{\bf r'}|} d{\bf r} d{\bf r'} \approx 0.34,
\end{equation}

where the ``inverse characteristic distance'' of $0.34$ is in units
determined by the lattice spacing. The Hartree-Fock results are
slightly sensitive to the value given here. However, when correlations
are taken into account the results are not sensitive to this value.

As a first approximation for the ground state of the Hamiltonian
(\ref{H}) we use the unrestricted and restricted Hartree-Fock (HF) methods,
modified according to the Optimal Damping Algorithm of Cances and Le
Bris\cite{cances00}.  Recall that the HF equations are of the form
\begin{equation}
\label{hfe}
H_{\sigma_i} \psi_i = \epsilon_i \psi_i \ ,
\end{equation}
where $\psi_i(r)$ is a single electron orbital which has energy
$\epsilon_i$, and spin projection on the z-axis $\sigma_i=\pm \frac{1}{2}$.  The
HF Hamiltonian matrix is of the form
\begin{equation}
\langle r|H_{\sigma_i}|r'\rangle = \langle r|H_s|r'\rangle + \langle r|U_{dir}|r'\rangle - 
\langle r|U_{exch, \sigma_i}|r'\rangle \ ,
\label{eqn:HFmatrix}
\end{equation}
where 
\begin{eqnarray}
\label{HFm1}
&&\langle r|U_{dir}|r'\rangle= \delta_{r r'} \ q^2 \sum_{\epsilon_k\le \epsilon_F} \sum_{r''} 
\frac{\psi^{\dag}_k(\mathbf{r''}) \psi_k(\mathbf{r''})}{|\mathbf{r} - \mathbf{r''}|} \ ,\nonumber
\\
&&\langle r|U_{exch, \sigma_i}|r'\rangle =q^2 \sum_{\epsilon_k \le \epsilon_F} \delta_{\sigma_i
\sigma_k} \frac{\psi^{\dag}_k(\mathbf{r'})
\psi_k(\mathbf{r})}{|\mathbf{r} - \mathbf{r'}|} \ ,
\end{eqnarray}
are direct and exchange interactions.  The single-particle Hamiltonian
has only diagonal and nearest neighbour nonzero matrix elements,
\begin{equation}
\label{sp}
\langle r|H_s|r'\rangle=\delta_{rr'} \ \left(4+\frac{1}{4}
(\overline{\omega}_x^2 x^2+ \overline{\omega}_y^2 y^2)\right) -\delta_{<rr'>} \ .
\end{equation}
Hear $\delta_{rr'}$ is the usual Kronecker delta.
$\delta_{<rr'>}=1$ if $r$ and $r'$ are nearest neighbours and
$\delta_{<rr'>}=0$ otherwise.  The total energy of the system is
\begin{widetext}
\begin{eqnarray}
\label{eqn:HFen}
E_{\textrm{HF}}  = \sum_{\epsilon_i \le \epsilon_F}  \Big( \langle H_s \rangle_i +
\frac{1}{2}\langle U_{dir} \rangle_i - \frac{1}{2}\langle U_{exch} \rangle_i \Big) 
=\sum_{\epsilon_i \le \epsilon_F}  \Big( \epsilon_i -
\frac{1}{2}\langle U_{dir} \rangle_i + \frac{1}{2}\langle U_{exch} \rangle_i \Big) \ .
\end{eqnarray}
\end{widetext}

In the unrestricted HF (UHF) method the single particle orbitals with
spin up are completely independent of those with spin down. Therefore,
generally UHF spontaneously violates the rotational invariance of the
Hamiltonian (\ref{H}). The total spin ${S}$ does not commute with the
UHF Hamiltonian and hence only the z-projection of the total spin, $S_z$, is
conserved. To characterise this effect we introduce the parameter
``spin separation''
\begin{equation}
\label{ssep}
\Delta n = \frac{1}{2}\int \big| n_\uparrow(\mathbf{r}) -
n_\downarrow(\mathbf{r}) \big| d\mathbf{r} \ .
\end{equation}
Here $n_\uparrow$ is the total density of electrons with spin up and
$n_\downarrow$ is the total density of electrons with spin down.  The
spin separation measures the spatial separation of up and down
electrons.  It can be nonzero even for a state with $S_z=0$.  Strictly
speaking the spontaneous violation of rotational symmetry and hence
nonzero value of $\Delta n$ for a state with $S_z=0$ is not a physical
effect, but a byproduct of the method. Nevertheless, a calculation
of $\Delta n$ can shed light on the physics of the system. For example,
zero $\Delta n$ indicates that there is no spontaneous symmetry
violation. If $\Delta n \approx 1$ in the sector with $S_z=0$ 
there are two possibilities: either the total spin is $S=1$ and we
effectively see the state $|S=1, S_z=0\rangle$, or the state is $|S=0,
S_z=0\rangle$. By performing the UHF calculation in the sector with
$S_z=1$ one can try to distinguish between these possibilities.

In the restricted HF (RHF) method the HF equations are exactly the
same, but in addition we impose the constraint that electron orbitals
with spin up and spin down are identical. Hence RHF automatically
respects the spin rotational invariance and describes the state with
$S=0$. The RHF total energy is always higher than the UHF total
energy.

\subsection{Correlation corrections}

\begin{figure}
\begin{center}
\includegraphics[width=0.35\textwidth]{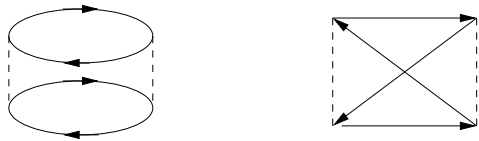}
\end{center}
\caption{The leading correlation correction to total energy}
\label{corr1}
\end{figure}

The leading correlation correction to the total energy is given by
the diagrams in Fig. \ref{corr1}.
The corresponding formula reads
\begin{widetext}

\begin{equation}
\label{c1}
\delta E_{\textrm{corr}}=\frac{1}{2}\sum_{\epsilon_i,\epsilon_j \le \epsilon_F}\sum_{\epsilon_m,\epsilon_n >\epsilon_F}
\frac{|\langle m,n|V|i,j\rangle|^2-\langle i,j|V|m,n\rangle\langle m,n|V|j,i\rangle}
{\epsilon_i+\epsilon_j-\epsilon_m-\epsilon_n} \ ,
\end{equation}

where the Coulomb matrix element is defined as

\begin{equation}
\langle m,n|V|i,j\rangle=\delta_{\sigma_m\sigma_i}\delta_{\sigma_n\sigma_j}  
\sum_{\bf r,r'}\psi_m({\bf r})\psi_n({\bf r'})
\frac{q^2}{|{\bf r}-{\bf r'}|}\psi_i({\bf r})\psi_j({\bf r'}) \ .
\label{matrixel}
\end{equation}
\end{widetext}

It is well known that in a 3D uniform electron gas the direct diagram
in Fig. \ref{corr1} is logarithmically divergent due to the long-range
nature of the Coulomb interaction. Hence it requires consideration of
screening effects (higher orders of perturbation theory) even at very
small $r_s$.\cite{GB} Fortunately for the 2D case the diagram is
convergent.  Moreover we consider a relatively small dot and in this
case the second order correlation diagram is convergent even in the 3D
case.

\subsection{Restricted HF with an odd number of electrons}
\label{sec:odd}

There is no difference between an even and an odd number of electrons
when performing UHF calculations. However, a restricted Hartree-Fock
calculation with an odd number of electrons is not straightforward as
the spin-up and spin-down electrons must feel a different exchange
potential, breaking the symmetry required by RHF.  The Slater
determinant of the problem consists of $(N-1)/2$ doubly occupied
orbitals which represent a core with total spin zero, and
also one upper orbital which contains only one electron. To preserve
the rotational invariance of the core we use the same method used in
atomic physics; we average over polarizations of the unpaired
electron.  This means that we introduce occupation numbers $n_i$ where
$n_i=1$ for any core orbital, $n_{e\uparrow}=n_{e\downarrow}=\frac{1}{2}$
where $e$ is the external orbital, and $n_i=0$ otherwise. The Hartree-Fock
equations (\ref{hfe})-(\ref{eqn:HFen}) are modified appropriately. For
example instead of Eq. (\ref{eqn:HFen}), the total energy reads
\begin{equation}
\label{eHF1/2}
E_{\textrm{RHF}} =\sum_{i}  \Big( n_i \epsilon_i -
\frac{1}{2} n_i \langle U_{dir} \rangle_i + \frac{1}{2} n_i \langle U_{exch}
\rangle_i \Big) \ .
\end{equation}
This procedure defines our first approximation and it corresponds to the many-body 
state with total spin $S=\frac{1}{2}$.

The next step is to take into account the fact that the real state
contains one $|e\uparrow\rangle$ electron and no $|e\downarrow\rangle$
electron. We use perturbation theory, which automatically guarantees
that the total spin remains $S=\frac{1}{2}$.  The transition from two
external ``half-electrons'' to one ``whole'' external electron results
in a perturbation acting on the external electron $\Delta
V_{e\uparrow} = -\frac{1}{2} \textrm{Dir}(e)$, and 
perturbations acting on the core electrons $\Delta V_{i\uparrow} =
-\frac{1}{2} \textrm{Ex}(e)$ and $\Delta V_{i\downarrow} = \frac{1}{2}
\textrm{Ex}(e)$.  Here $\textrm{Dir}(e)$ and $\textrm{Ex}(e)$
represent direct and exchange interaction with the ``external''
orbital $\psi_e$.  Applying these perturbations one finds the
following first- and second-order corrections to total energy
(\ref{eHF1/2}).
\begin{eqnarray}
\label{1/2}
\delta E^{(1)} & = & -\frac{1}{4} \langle e, e | V | e, e \rangle \ , \nonumber\\
\delta E^{(2)} & = & \frac{1}{4} \sum_{k>e} \frac{|\langle e, e | V | k, e
  \rangle|^2}{\epsilon_k - \epsilon_e}
+ \frac{1}{4} \sum_{i<e} \sum_{k>e} \frac{|\langle i, e | V | e, k
  \rangle|^2}{\epsilon_k - \epsilon_i}
+ \frac{1}{4} \sum_{i<e} \sum_{k\ge e} \frac{|\langle i, e | V | e, k
  \rangle|^2}{\epsilon_k - \epsilon_i} \ .
\end{eqnarray}
We use the notation defined in Eq. (\ref{matrixel}).
Note that (\ref{1/2}) is not the same as the correlation correction.
The correlation energy has to be calculated separately using (\ref{c1}).
In (\ref{1/2}) we go to second order in order to retain formal
accuracy comparable to the correlation terms (\ref{c1}).
However, in practice we have found that $\delta E^{(2)}$ is always 
very small and can be neglected.

\section{Two particles in a circular parabolic potential}
\label{sec:twobody}

\begin{figure}
\begin{center}
\includegraphics[width=0.27\textwidth]{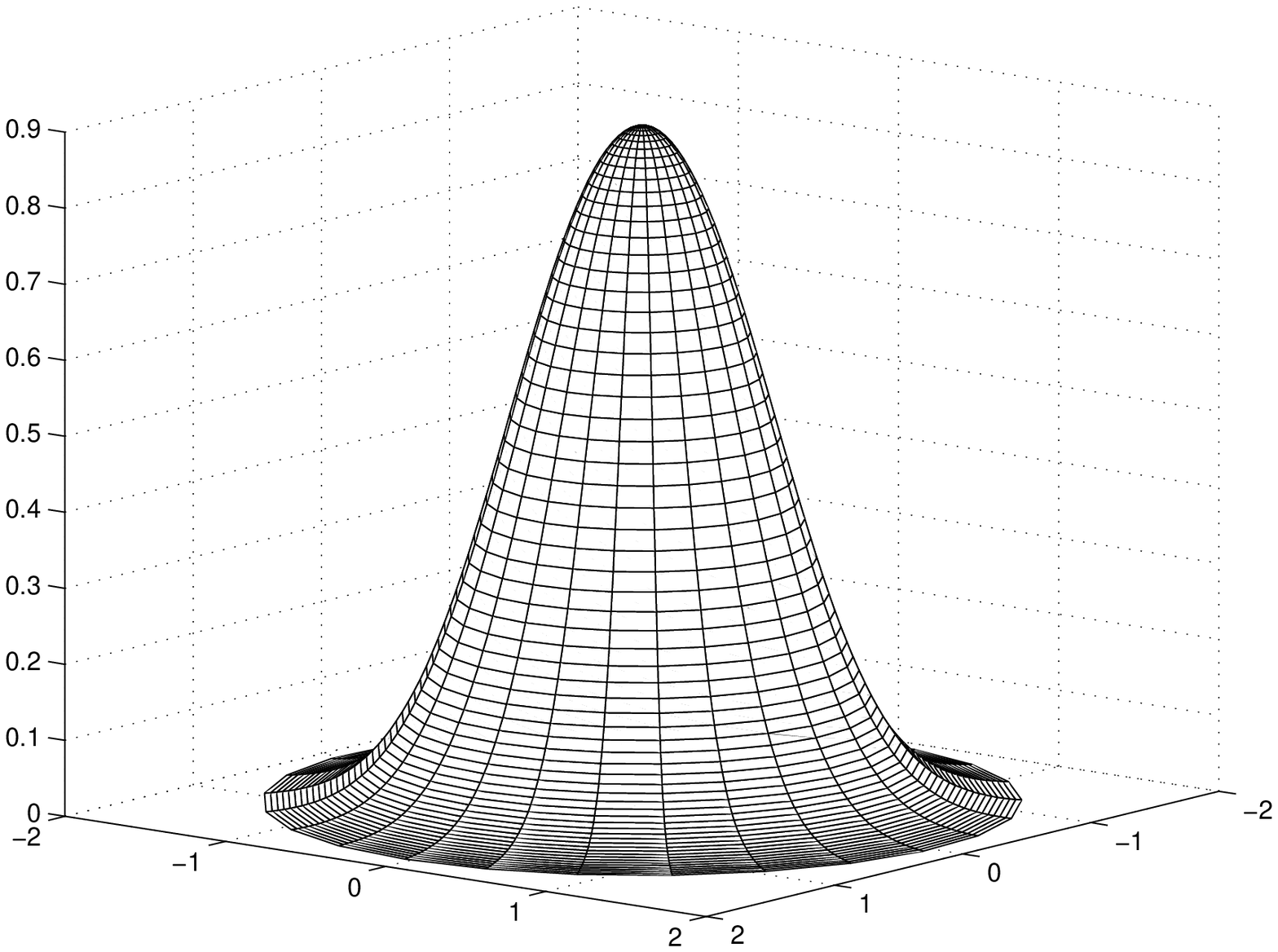}%
\hspace{0.5cm}%
\includegraphics[width=0.27\textwidth]{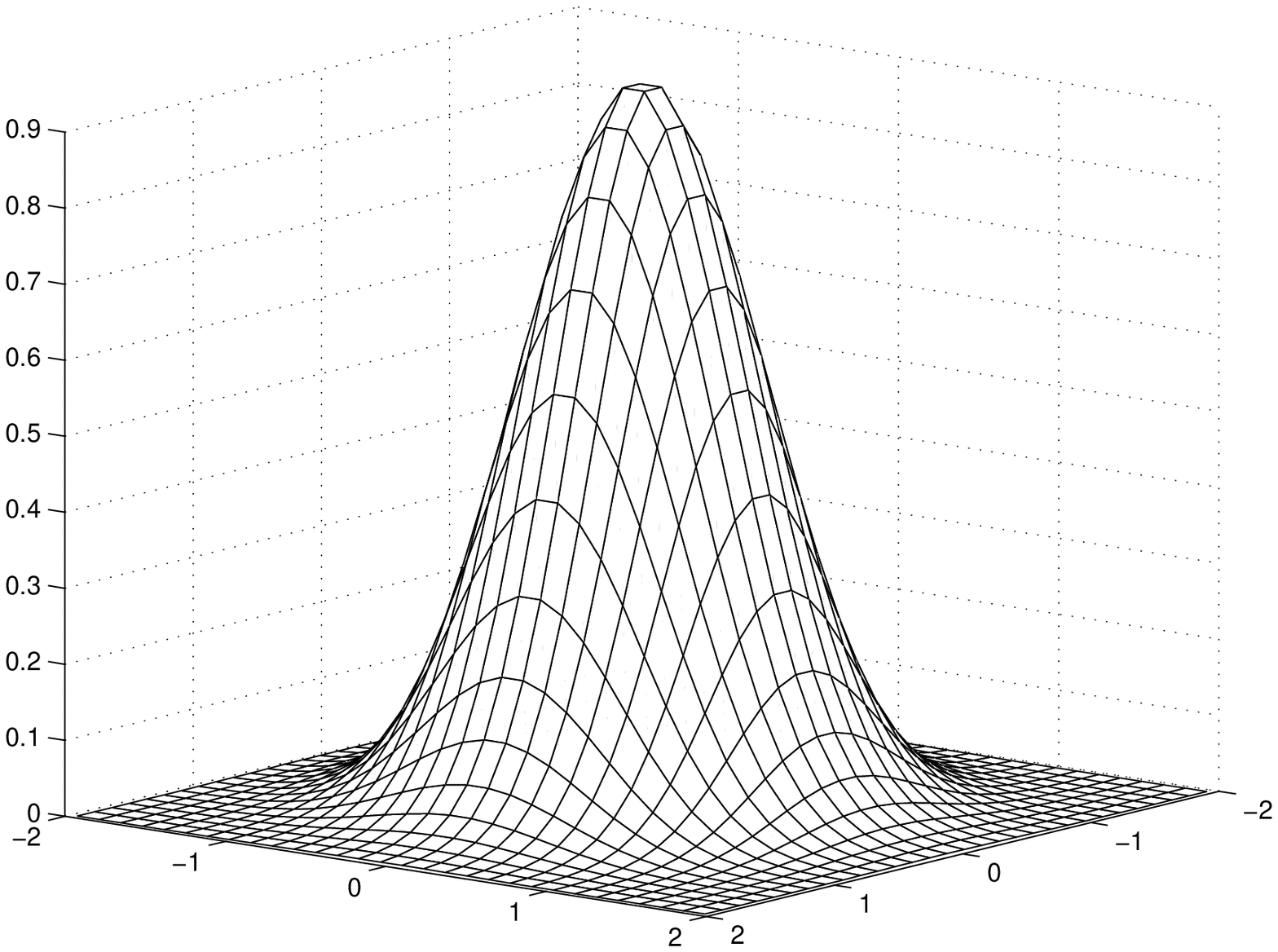}
\end{center}
\caption{Exact (left) and Hartree-Fock (right) charge densities
(atomic units) for a two-electron dot at $r_s = 0.77$.  In this case
restricted and unrestricted Hartree-Fock results are practically
identical. Note that $r_s$ here is given by the semiclassical estimate made in Section \ref{sec:estimates} and, for only two electrons, will not be exactly equal to $E_\textrm{Coulomb} / E_\textrm{Kinetic}$. }
\label{fig:compare0.5}
\end{figure}

\begin{figure}
\begin{center}
\includegraphics[width=0.27\textwidth]{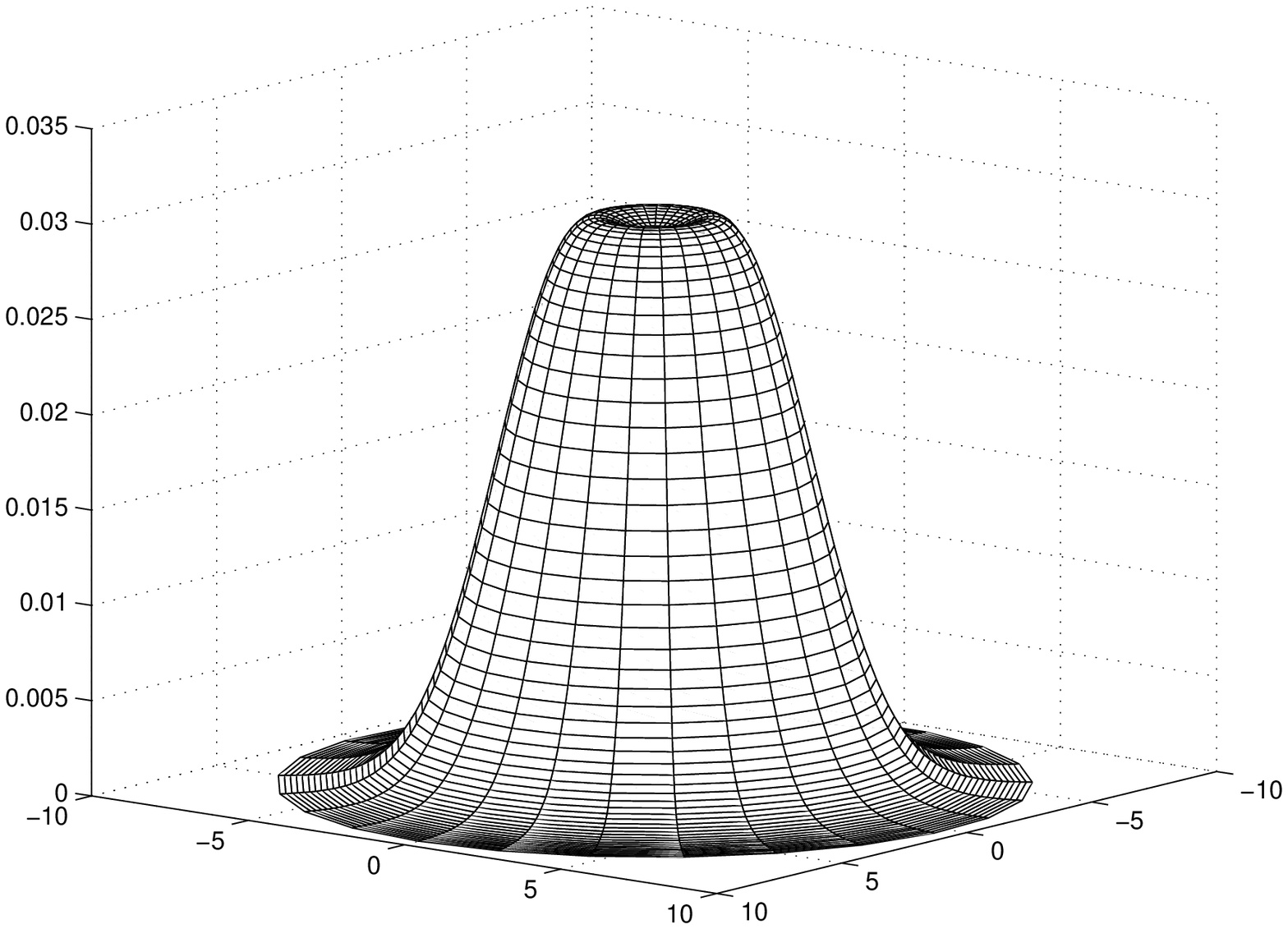}%
\hspace{0.5cm}%
\includegraphics[width=0.27\textwidth]{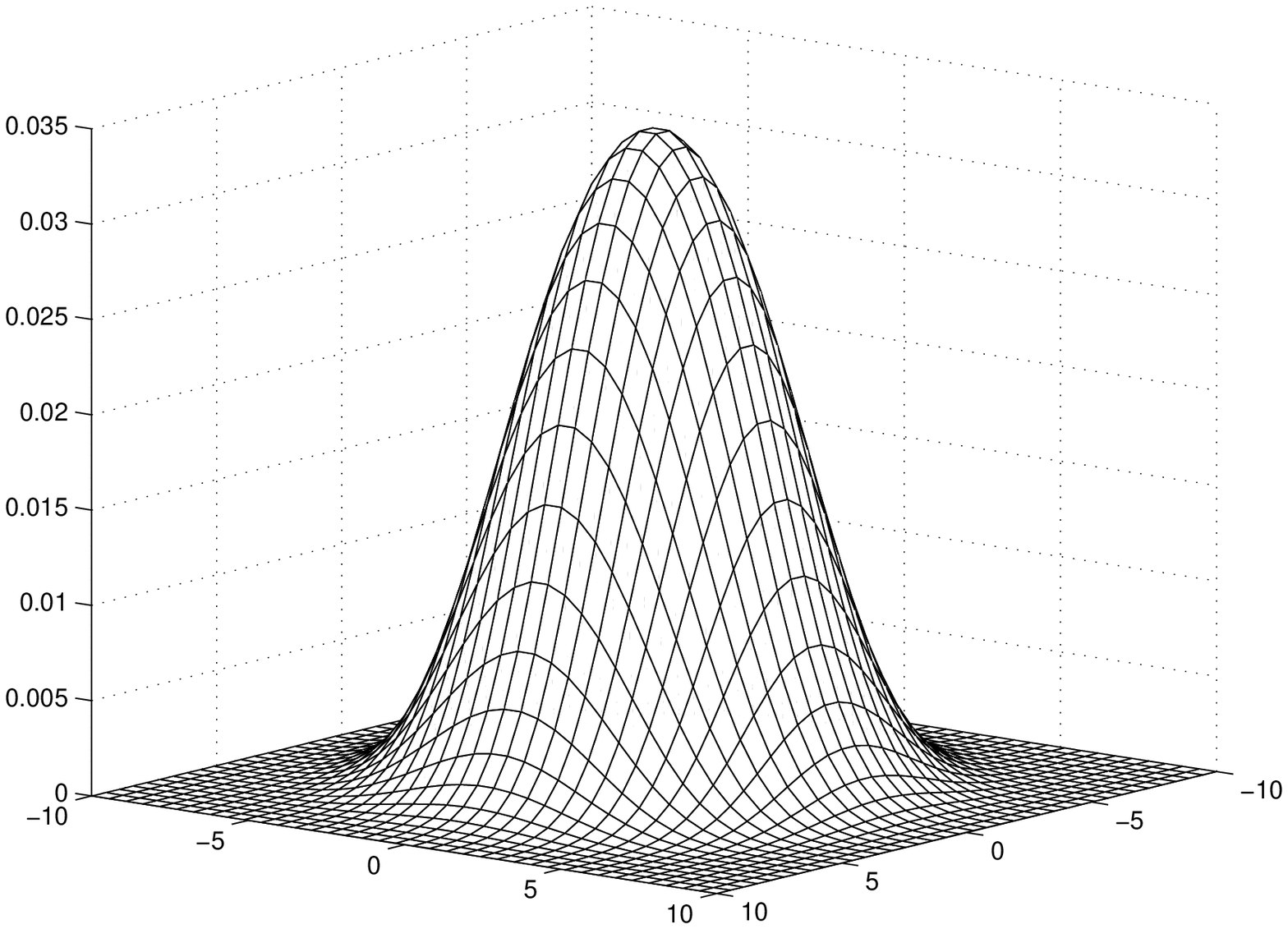}
\hspace{0.5cm}%
\includegraphics[width=0.27\textwidth]{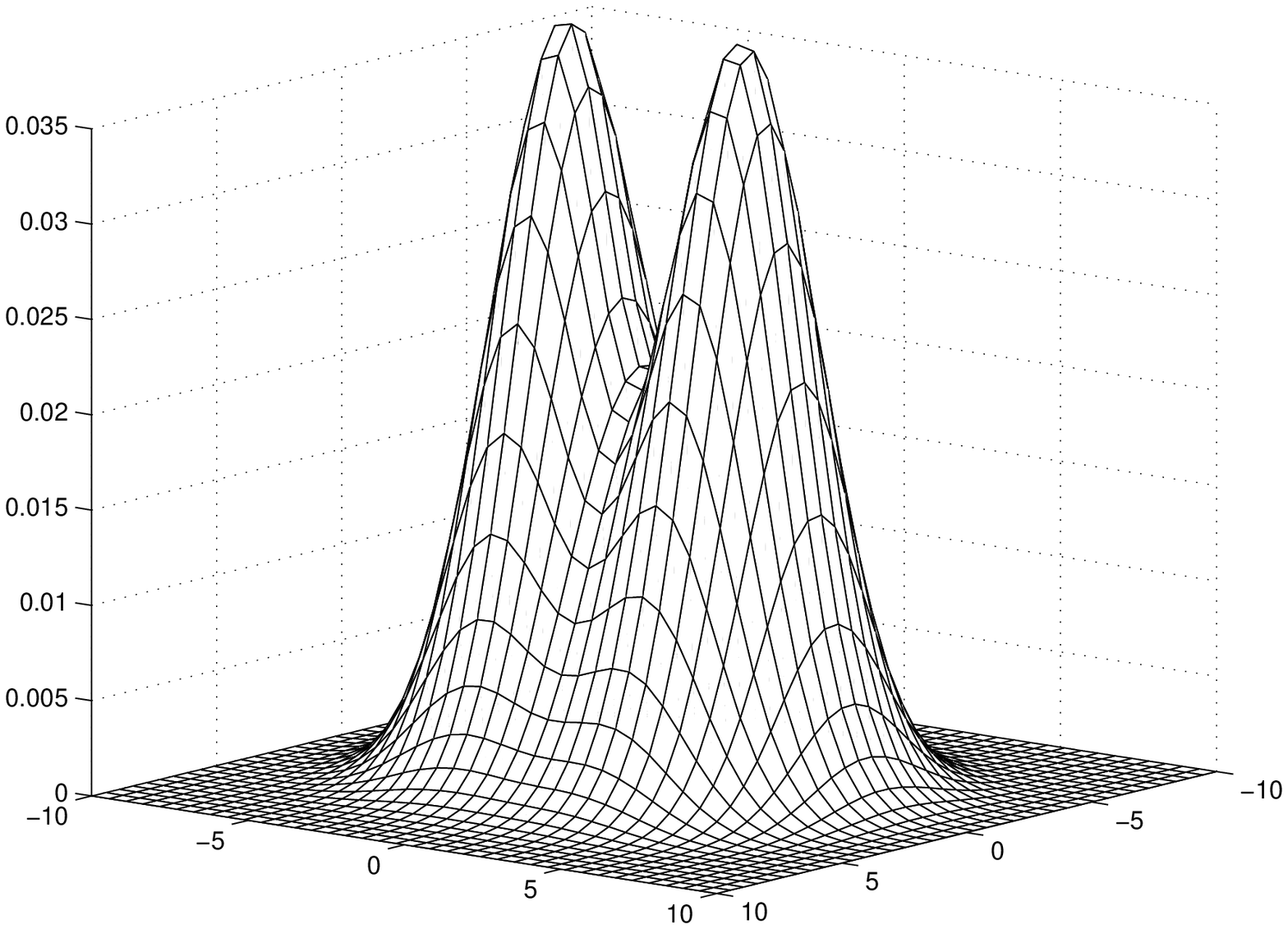}
\end{center}
\caption{Exact (left), restricted Hartree-Fock (central), and
unrestricted Hartree-Fock (right) charge densities (atomic units) for
a two-electron dot at $r_s = 4.87$. Note that $r_s$ here is given by the semiclassical estimate made in Section \ref{sec:estimates} and, for only two electrons, will not be exactly equal to $E_\textrm{Coulomb} / E_\textrm{Kinetic}$.}
\label{fig:compare2}
\end{figure}

\begin{figure}
\begin{center}
\includegraphics[width=0.4\textwidth]{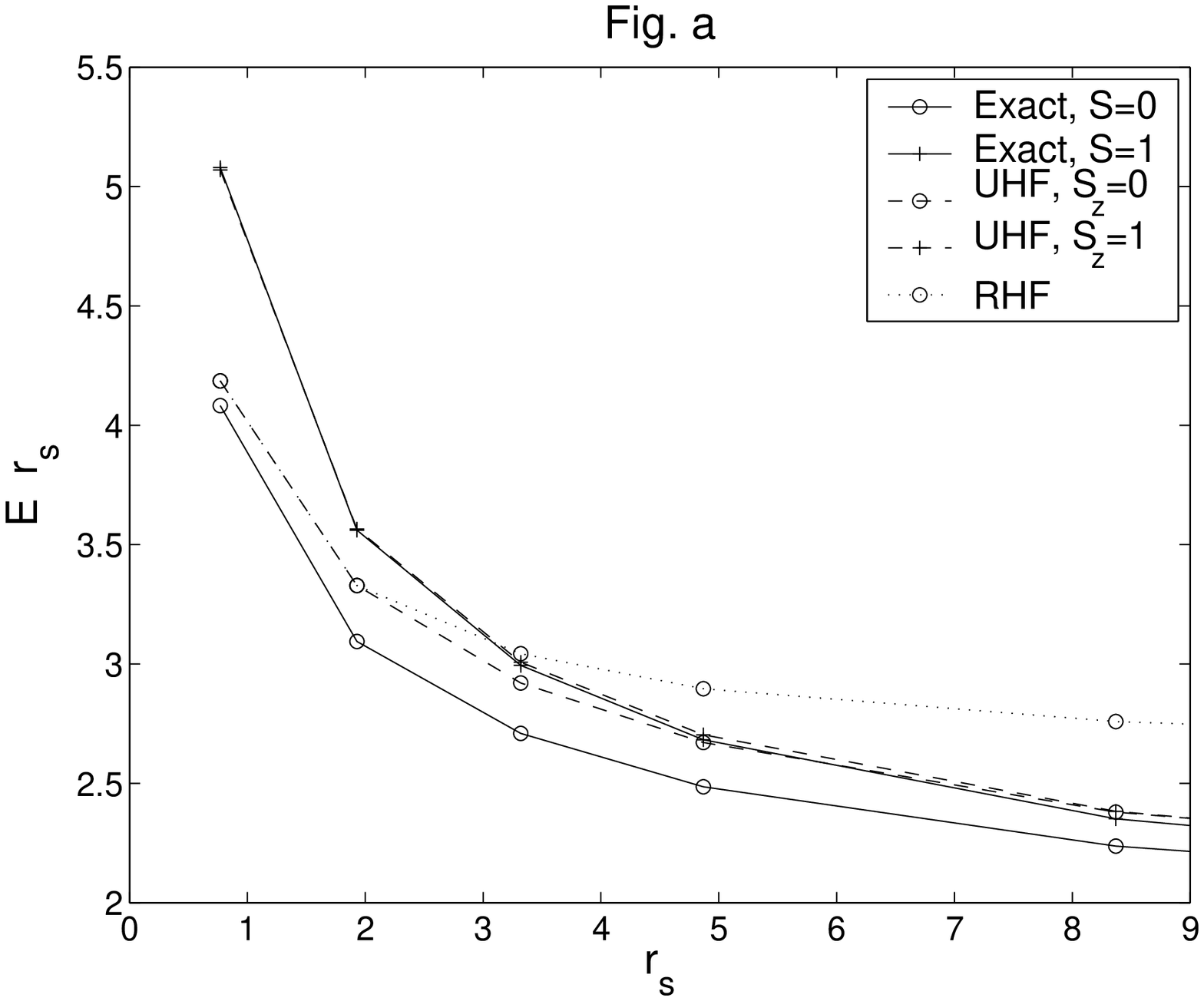}
\hspace{1.cm}
\includegraphics[width=0.4\textwidth]{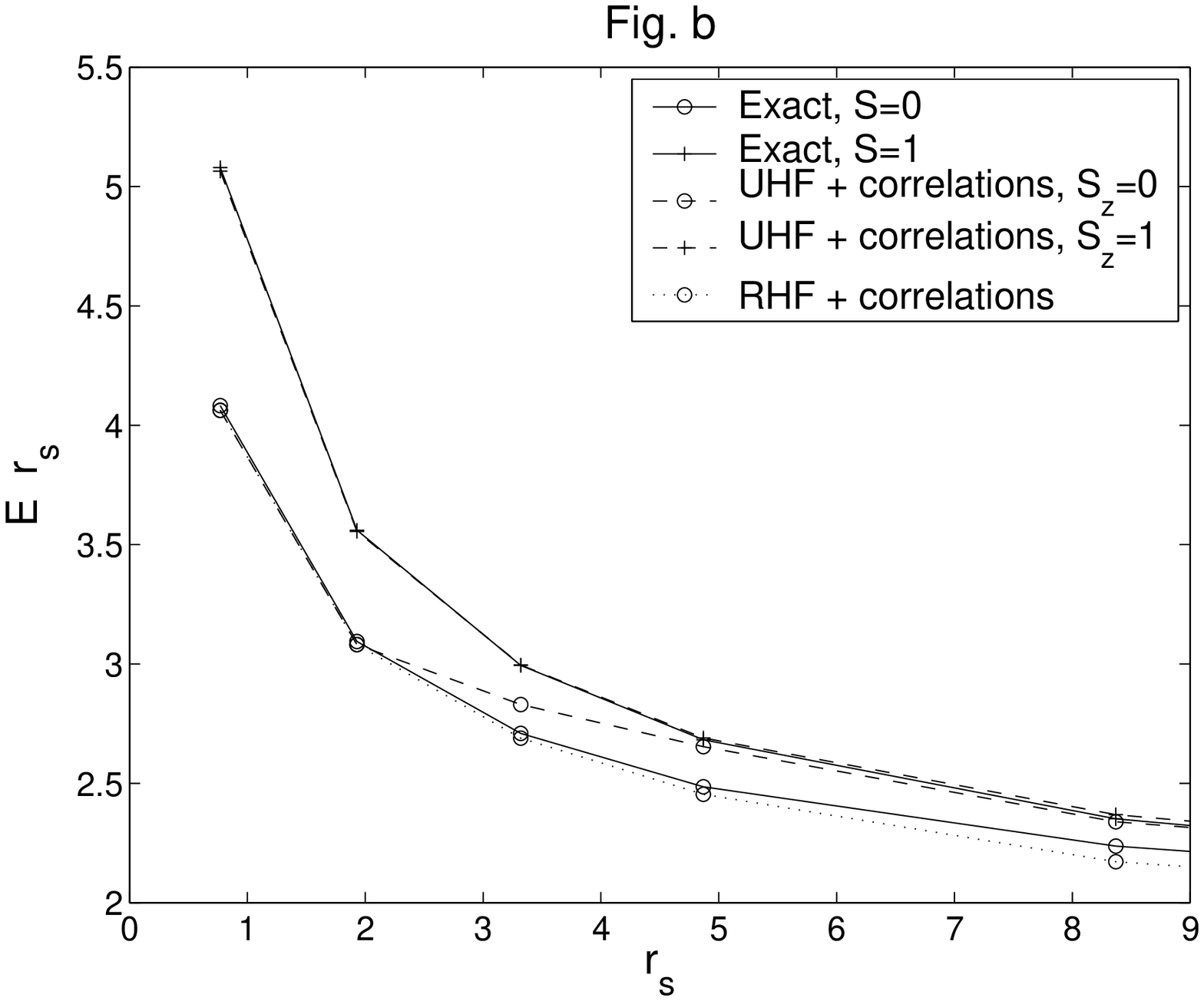}
\end{center}
\caption{Total energy versus $r_s$ for the two particle problem. We
plot the total energy in atomic units multiplied by $r_s$.  Exact
solutions for $S=0$ and $S=1$ are shown in both Fig.a and Fig.b by
solid lines.  Fig.a also shows the UHF energy with $S_z=0$ and $S_z=1$
(the dashed lines) and the RHF energy (the dotted line). Fig.b shows
the same energies with second-order correlation terms (\ref{c1})
included in all cases. Lines joining data points are given as a guide
to the eye only.}
\label{fig:twoelec}
\end{figure}

The two-body problem can be solved exactly. The Hamiltonian of the
problem is
\begin{equation}
H = \frac{\mathbf{p_1^2}}{2m}+\frac{\mathbf{p_2^2}}{2m}+
\frac{1}{2} m \omega^2 \big( 
| \mathbf{r_1} |^2 + | \mathbf{r_2} |^2 \big) 
+ \frac{e^2}{ | \mathbf{r_1} - \mathbf{r_2} | }
\end{equation}
Using a centre-of-mass coordinate and a relative position
coordinate:
\begin{eqnarray}
\mathbf{R} & = & \frac{1}{2}(\mathbf{r_1} + \mathbf{r_2})\ , \nonumber \\
\mathbf{r} & = & \mathbf{r_1} - \mathbf{r_2} 
\end{eqnarray}
one can transform the  Hamiltonian to
\begin{equation}
H = -\frac{1}{4m}\frac{\partial^2}{\partial\mathbf{R}^2}+
m \omega^2  \mathbf{R}^2 - \frac{1}{m}\frac{\partial^2}{\partial\mathbf{r}^2}
+\frac{1}{4} m \omega^2 |\mathbf{r} |^2 + \frac{e^2}{ | \mathbf{r} | } \ .
\end{equation}
Variables in the corresponding Schroedinger equation are separated and
hence one can easily find the exact solution of the problem by solving
the radial equation for $\mathbf{r}$ numerically. The ground state has
relative orbital angular momentum equal to zero, $l=0$. Therefore,
because of Fermi statistics, the spin of the ground state is also
zero, $S=0$. The first excitation has $l=S=1$.  For clarity, in
Fig. \ref{fig:twoelec} we plot values of $E r_s$ versus $r_s$,
where $E$ is the energy in atomic units and $r_s$ is defined according
to (\ref{r1}) at $N=2$ (see comment \cite{com1}).

Fig. \ref{fig:twoelec}a shows exact energies for $S=0$ and $S=1$, UHF
energies for $S_z=0$ and $S_z=1$, and the RHF $S=0$
energy. Fig. \ref{fig:twoelec}b shows the same energies but with
second order correlation corrections included (\ref{c1}). In this case
the dashed lines show $E_\textrm{UHF}+\delta E_\textrm{UHF}$ at
$S_z=0$ and $S_z=1$.

In Fig. \ref{fig:compare0.5} and Fig. \ref{fig:compare2} we show
exact, RHF, and UHF electron densities in the ground state for
$r_s=0.77$ and $r_s=4.87$ respectively.  At small $r_s$ both the
restricted and unrestricted ($S_z=0$) Hartree-Fock methods give
results very close to the exact $S=0$ solution. However, at larger
$r_s$ the ground state UHF energy is lower than that obtained in the
restricted Hartree-Fock method (see Fig. \ref{fig:twoelec}a).  The UHF
energy for the ground state approaches the energy of the first
excitation with $S=1$ at larger values of $r_s$.

This illustrates that the UHF method spontaneously violates the spin
and rotational symmetry of the problem and mixes the true ground
state, which has $S=0$, with the state $|S=1,S_z=0\rangle$.  At $r_s >
2$ the $S=1$ component dominates. This leads to early ``Wigner
crystallization'' which is a byproduct of the UHF method. This is
confirmed by the pictures of electron density at $r_s= 4.87$ shown in
Fig. \ref{fig:compare2}. There are two separate peaks in the UHF
electron density; one corresponds to the electron with spin up and
another to the electron with spin down. The UHF spin separation
parameter (\ref{ssep}) is $\Delta n=1.7$. At the same time neither
the exact solution nor the RHF solution indicate any spin separation.

Taking correlations into account drastically changes the
situation. The total RHF+correlation energy shown by the dotted line
Fig. \ref{fig:twoelec}b is close to the exact energy up to very high
$r_s$ for the two-electron problem and is below the corresponding
energy obtained with the UHF+correlations method.  Thus the inclusion
of correlations leads to the choice of the RHF solution as the ground
state, hence restoring the rotational invariance at $S=0$. Certainly
at very large $r_s$, all Hartree-Fock methods fail.

\begin{figure}
\begin{center}
\includegraphics[width=0.4\textwidth]{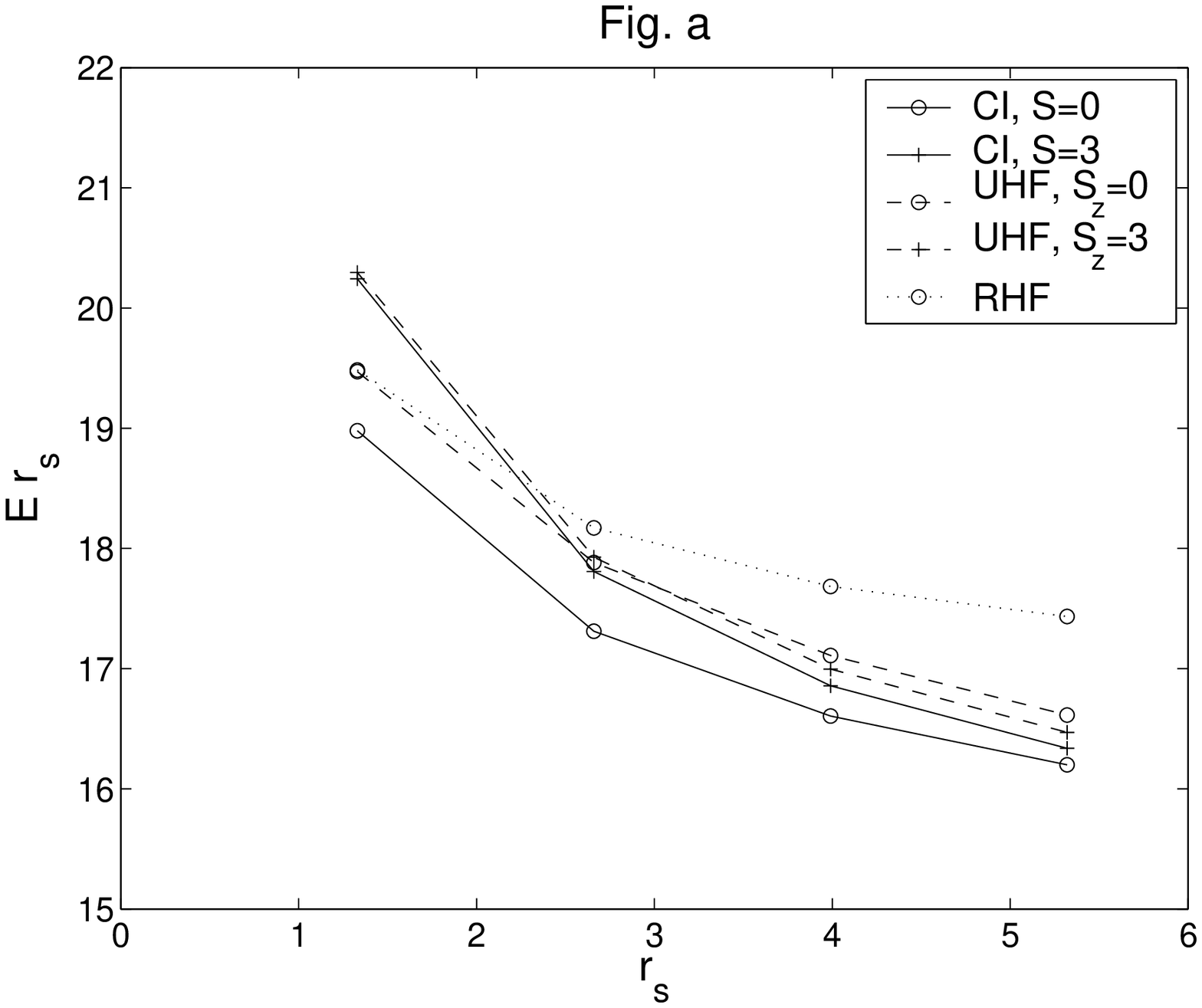}
\hspace{1.cm}
\includegraphics[width=0.4\textwidth]{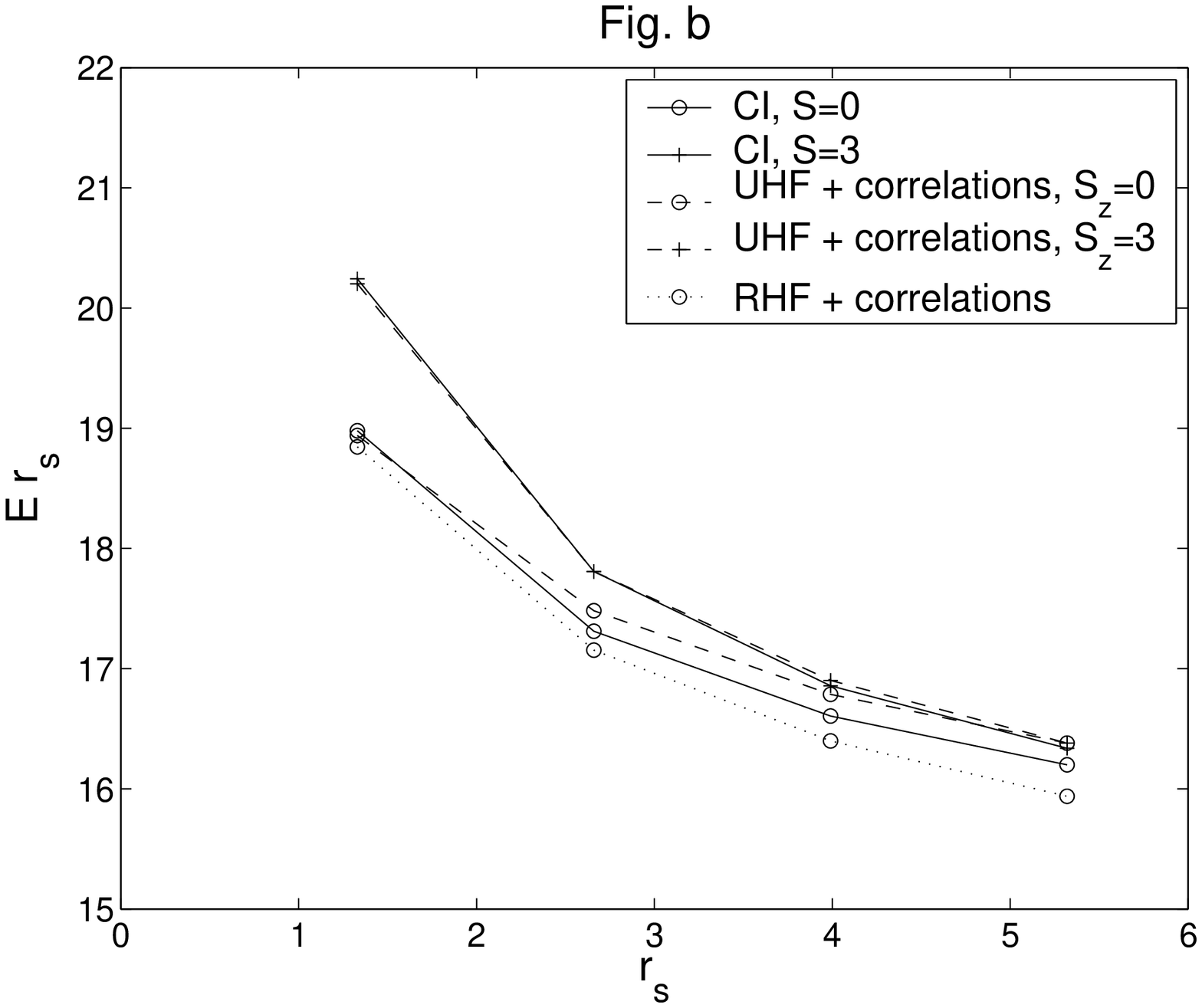}
\end{center}
\caption{Total energy versus $r_s$ for the six particle problem, with
CI energies taken from the configuration-interaction study of
Reimann \emph{et al.}\cite{Reimann00} We plot the total energy in
atomic units multiplied by $r_s$.  Exact solutions for $S=0$ and $S=3$
are shown in both Fig.a and Fig.b by solid lines.  Fig.a also shows
the UHF energy with $S_z=0$ and $S_z=3$ (the dashed lines) and the RHF
energy (the dotted line). Fig.b shows the same energies with
second-order correlation terms (\ref{c1}) included in all cases. Lines
joining data points are given as a guide to the eye only.}
\label{fig:sixelec}
\end{figure}

Fig. \ref{fig:sixelec} gives a similar comparison for the six-electron
dot, with configuration-interaction values taken from the 
 study of Reimann \emph{et al.}\cite{Reimann00}. Here
Fig. \ref{fig:sixelec}a shows CI energies for $S=0$ and $S=3$, UHF
energies for $S_z=0$ and $S_z=3$, and the RHF $S=0$ energy, while
Fig. \ref{fig:twoelec}b shows the same energies but with second order
correlation corrections included (\ref{c1}). We see the same behaviour
as in the two-electron problem, although the third and higher order
terms not included in our method are larger when N=6. Again the true
ground state is $S=0$, and as $r_s$ increases the UHF $S_z = 0$
solution breaks the symmetry of the problem and converges to the
polarised solution. By using RHF as a base for the perturbation theory
we remove the spin density waves and restore the rotational invariance
of the ground state.

We point out that the UHF and the UHF+correlations method both work
pretty well for the polarised $S=1$ and $S=3$ states, see
Fig. \ref{fig:twoelec}a,b.  This is because the coordinate wave
function is antisymmetric with respect to the permutation of electrons
and hence the electrons are well separated in space.

\section{The multi-electron circular quantum dot}
\label{sec:round}

\begin{figure}
\begin{center}
\includegraphics[width=.4\textwidth]{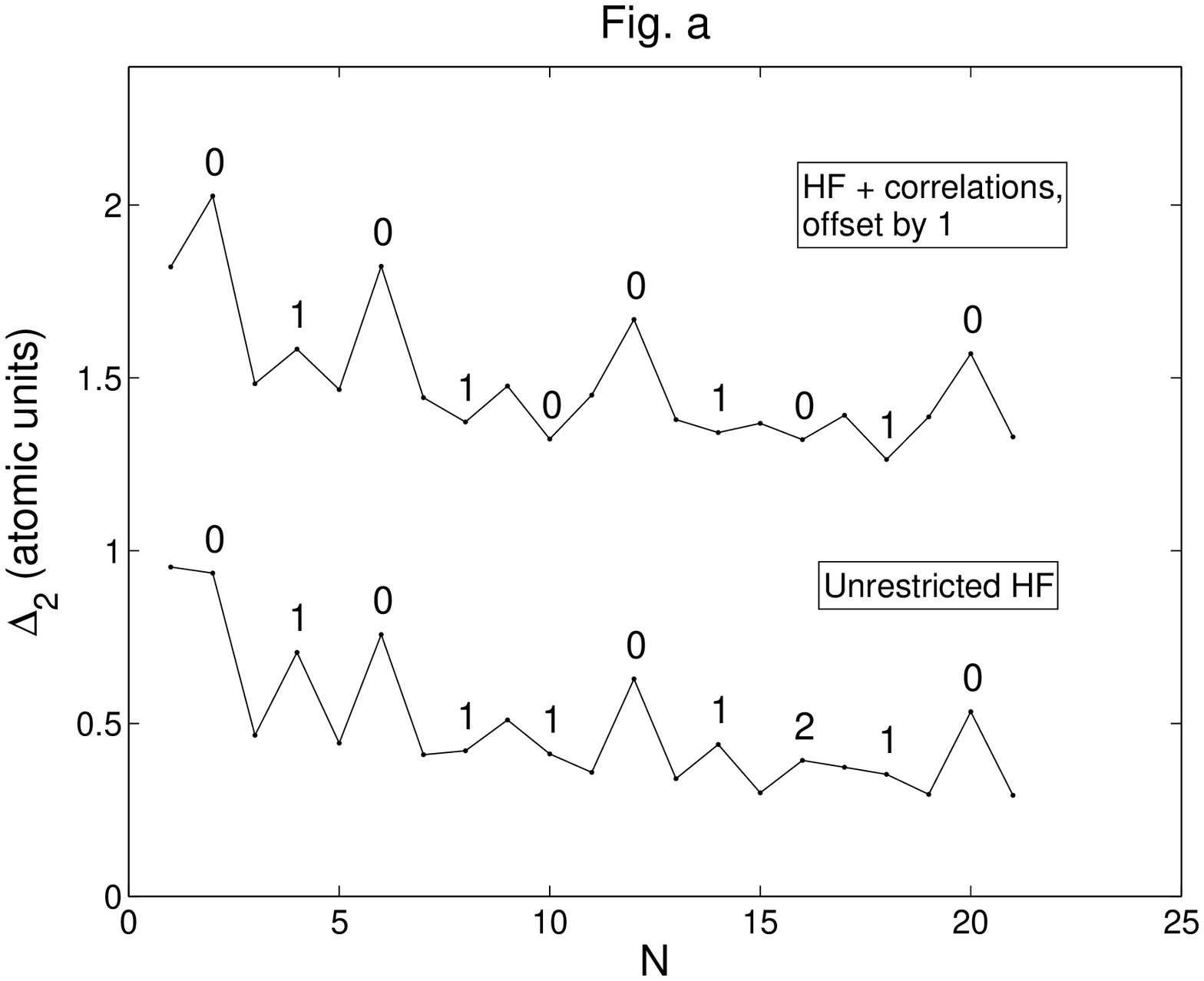}
\includegraphics[width=.4\textwidth]{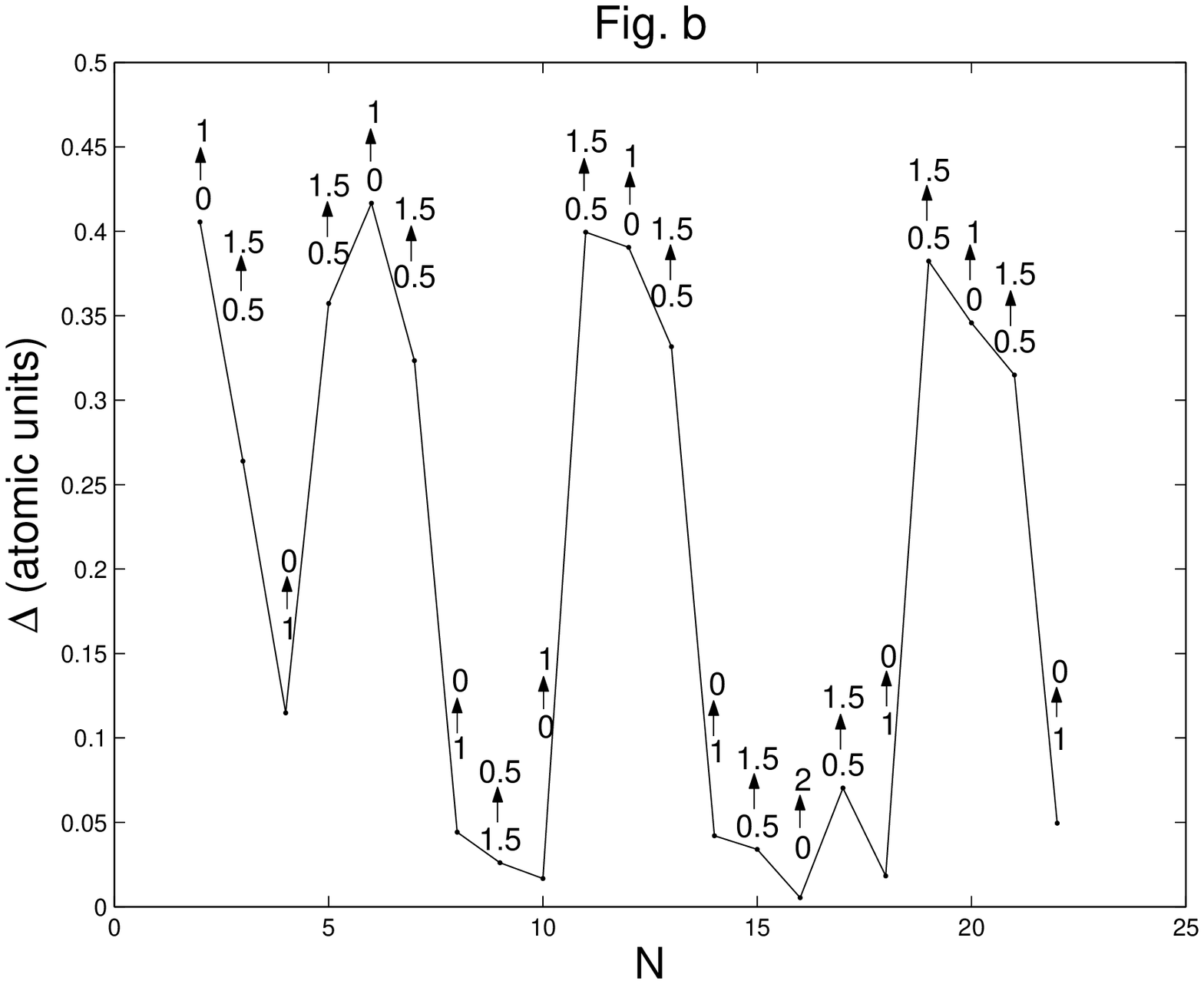}
\end{center}
\caption{The circular dot with confinement frequency $\omega_{\textrm{at}} =
0.73$ which corresponds to $r_s = 1.48 \ldots 0.99$ over the range of
$N$.\\
{\bf a}: $\Delta_2$ against $N$ from unrestricted HF (bottom line)
and from HF+correlations (top line). At even values of $N$ we also show
the spin of the ground state. \\
{\bf b}: The excitation gap with spin change against $N$ (correlations are taken into account). At each N we also show the spin of the ground state and the spin of the excited state.
}
\label{fig:w0.73}
\end{figure}

The second difference of the total energy is defined in the usual way
\begin{equation}
\Delta_2(N)=E(N+1)+E(N-1)-2E(N) \ ,
\end{equation}
where $N$ is the number of electrons in the dot.
Fig. \ref{fig:w0.73}a shows calculated values of $\Delta_2$ against
$N$ in the round dot with confinement frequency $\omega_{\textrm{at}}
= 0.73$ (atomic units) which corresponds to $r_s = 1.48 \ldots 0.99$
over the range of $N$ (see Eq. \ref{r1}).  We also find the ground
state spin. Values of the spin at even values of $N$ are marked above
the lines. The bottom line in Fig. \ref{fig:w0.73}a describes naive HF
calculations with no correlations included.  These are all UHF results
since the raw UHF energy will always be lower than the raw RHF energy.
To determine total spin of the UHF state we perform calculations for
different values of $S_z$ and set $S$ to the highest of the $S_z$
which give degenerate energies. In cases when $S \ne 0$, we see very
degenerate energies up to the maximal spin, followed by a much larger
energy gap for $S_z>S$, and this indicates that UHF almost preserves
the rotational invariance. These UHF results agree with those obtained
previously using spin density functional theory
\cite{KMR,Reimann}. The 2D parabolic shell structure is reflected in
the $\Delta_2$ values. We see peaks at $N = 2$, $6$, $12$, and $20$,
consistent with the shell model $1s^2 2p^4 3s^2 3d^4 4p^4 4f^4$. As
one would expect, the Hartree-Fock spin structure agrees with that
predicted by Hund's rule.  
\begin{figure}
\begin{center}
\includegraphics[width=.4\textwidth]{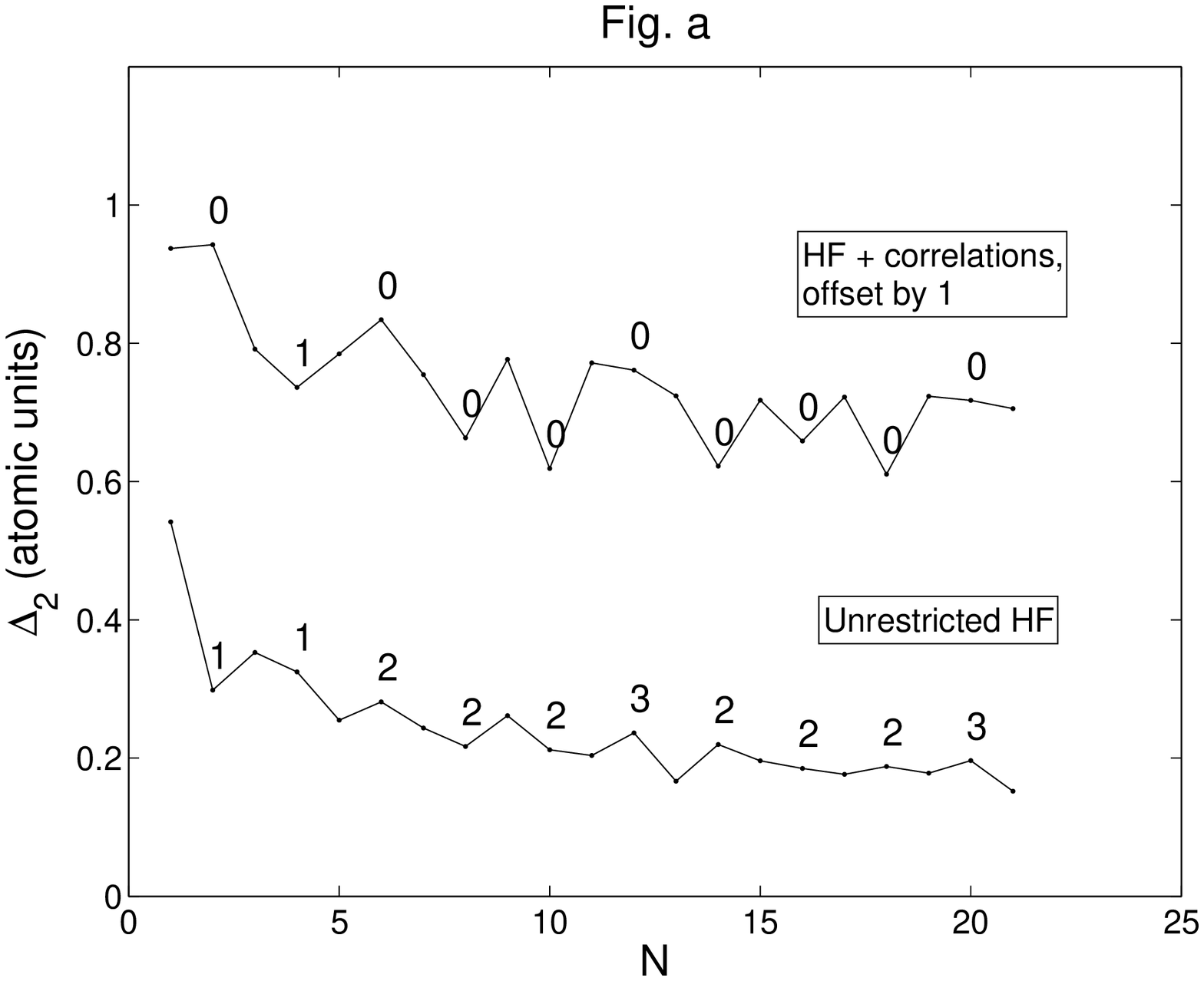}
\includegraphics[width=.4\textwidth]{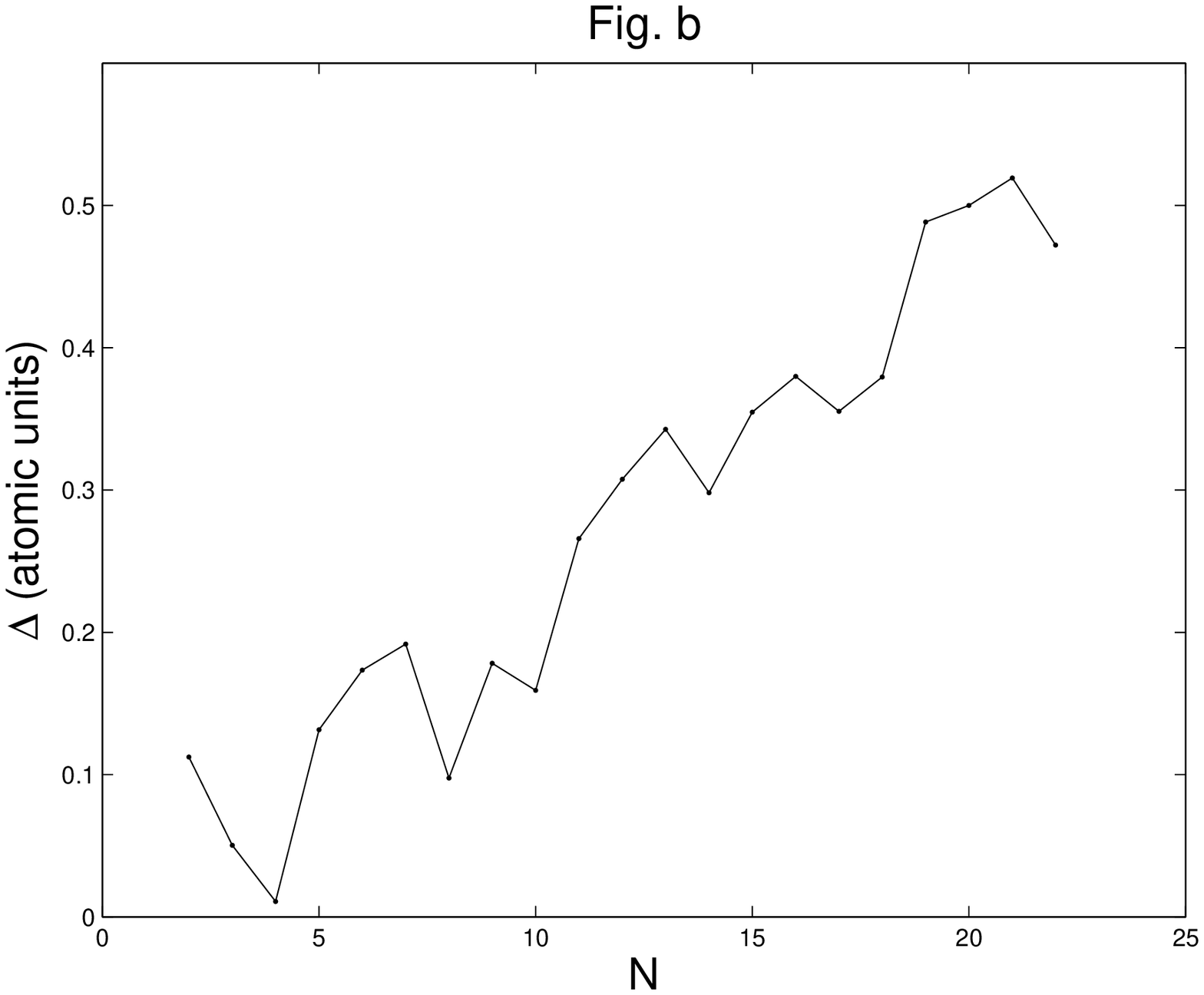}
\end{center}
\caption{The circular dot with confinement frequency $\omega_{\textrm{at}} =
0.26$ which corresponds to $r_s = 2.91 \ldots 1.95$ over the range of
$N$.\\
{\bf a}: $\Delta_2$ against $N$ from unrestricted HF (bottom line)
and from HF+correlations (top line). At even values of $N$ we also show
the spin of the ground state. \\
{\bf b}: The excitation gap against $N$ as defined by $E_{\textrm{UHF, Sz =1}} - E_\textrm{RHF, S=0}$ (correlations are taken into account).
}
\label{fig:w0.26}
\end{figure}

The top line (offset) in Fig. \ref{fig:w0.73}a shows results obtained
by taking correlations into account.  At each $N$ we perform
UHF+correlation calculations for different values of $S_z$ and we also
perform a RHF+correlation calculation for $S=0$ (or $S=\frac{1}{2}$ if
$N$ is odd). This allows us to determine the ground state energy and
spin. Fig. \ref{fig:w0.73}a shows that correlations influence the
values of $\Delta_2$ for some $N$, but retain the shell structure. For
some $N$ where the shell is open, i.e. $N=10$, $15$, $16$, $17$, the
correlations also change the spin from $1$, $\frac{3}{2}$, $2$,
$\frac{3}{2}$ to $0$, $\frac{1}{2}$, $0$, $\frac{1}{2}$. In these
cases, including correlations leads to a violation of Hund's rule.
However, in these cases we see that the excitation gap with change of
spin
\begin{equation}
\label{gapp}
\Delta = E_1(N)-E_0(N)
\end{equation}
is very small in these dots. The spin gap becomes large only for $N$
corresponding to closed shells. Values of $\Delta$ versus $N$ at
$\omega_{\textrm{at}} = 0.73$, with correlations taken into account,
are plotted in Fig. \ref{fig:w0.73}b.  Unfortunately our technique
does not allow the calculation of the gap without spin change.  Note
that this gap is \emph{not} equal to the difference of the
single particle Hartree-Fock energies. We should point out that at
$N=16$, where the calculated excitation goes from $S=0$ to $S=2$, we
cannot accurately calculate the energy at $S=1$ as only $S_z$ is
conserved in the HF method. This means we cannot observe the $S=1$
excitation - most likely it has energy comparable to that of the $S=2$
excitation.

Fig. \ref{fig:w0.26} shows the same quantities $\Delta_2$ and $\Delta$
for dots with weaker confinement $\omega_{\textrm{at}} = 0.26$. This
corresponds to stronger interaction, $r_s = 2.91 \ldots 1.95$ over the
range of $N$. In this case correlations substantially influence the
values of $\Delta_2$, although the shell structure on $\Delta_2$ is
destroyed by electron-electron interactions whether or not
correlations are included.  It should be noted that in the UHF case
here it becomes difficult to determine which energies should be
treated as degenerate and which should not. The UHF spin values given
on the bottom line of Fig. \ref{fig:w0.26}a should be taken with
caution. This is in contrast to Fig. \ref{fig:w0.73}a where the
degeneracies are clear-cut.  When correlations are included, the
ground state spin becomes minimal for almost all $N$.  We see in Fig
\ref{fig:w0.26}b that the ``spin gap'' of Eq. (\ref{gapp}) grows linearly
with $N$.  At the same time, the UHF+correlations $S_z=1$ calculation
in this case generates a highly excited state with a strong spin
density wave (the spin separation parameter is $\Delta n \sim
4-5$). This implies that the accuracy of the UHF+correlations method
drops dramatically at the larger $N$ due to the developement of
unphysical spin density waves, and that the gap in
Fig. \ref{fig:w0.26}b is not a true spin gap. We believe that the
$S=0$ ground state, as determined by RHF+correlations, is fairly
accurate. In principle, based on this ground state, one could find the
first physical excitation using the time-dependent HF+correlations
method. However this is a very involved calculation and beyond the
scope of the present work.

\section{Elliptical dots}
\label{sec:elliptical}

\begin{figure}
\begin{center}
\includegraphics[width=.4\textwidth]{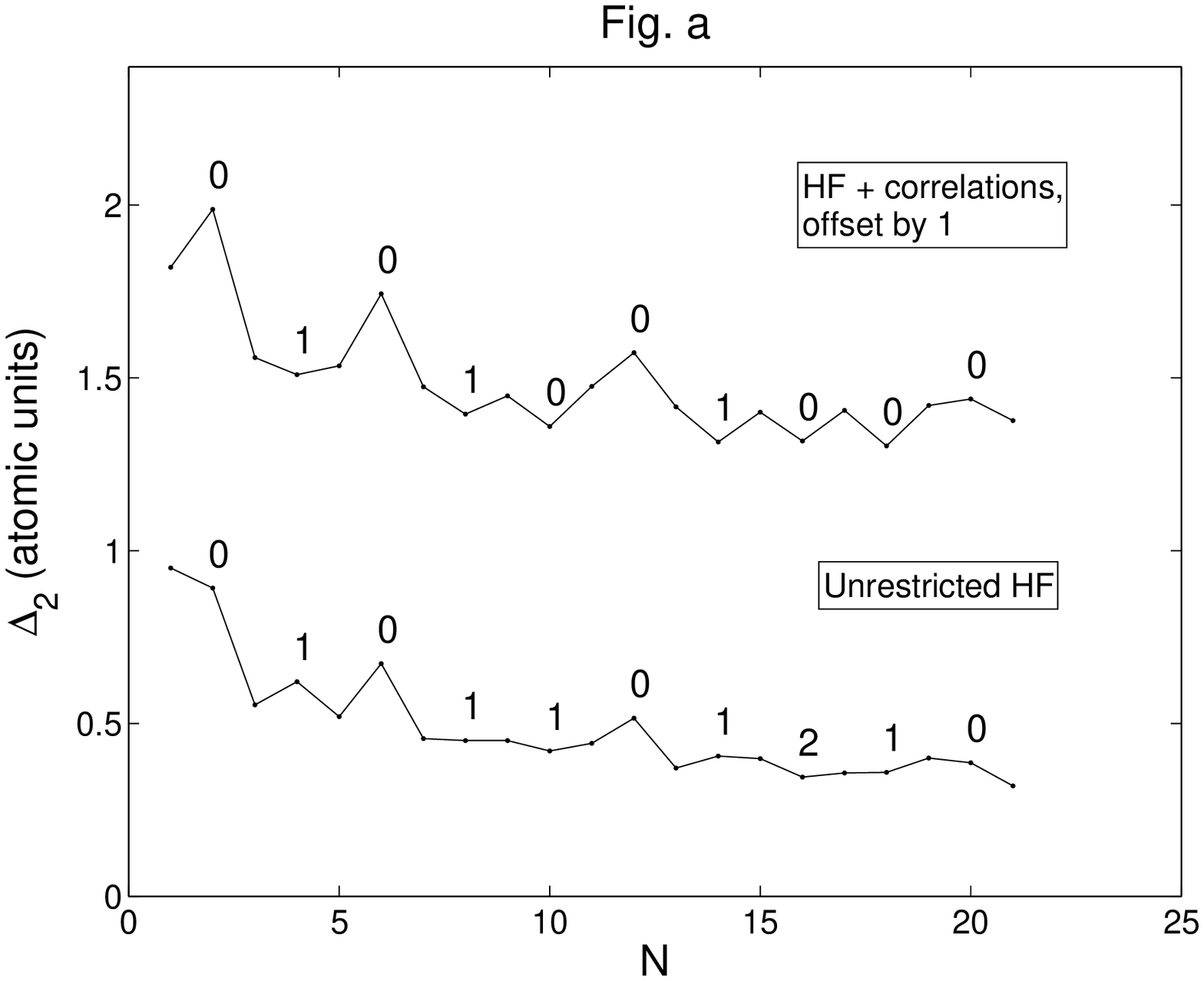}
\includegraphics[width=.4\textwidth]{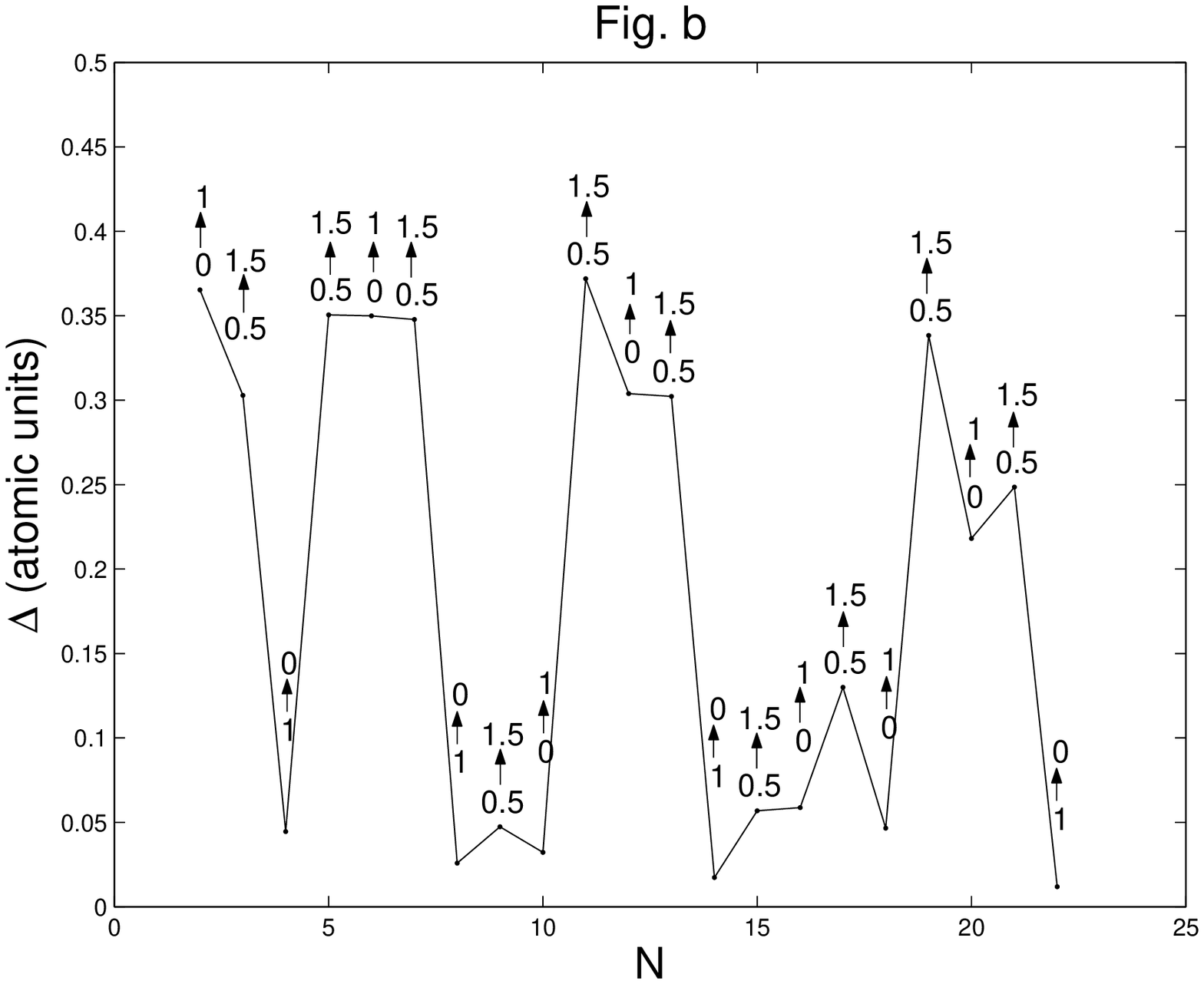}
\end{center}
\caption{The dot with ellipticity $\omega_x/\omega_y = 1.1$ and with
$\omega_{\textrm{at}}=\sqrt{\omega_x\omega_y}= 0.73$ which corresponds to $r_s
= 1.48 \ldots 0.99$.\\
{\bf a}: $\Delta_2$ against $N$ from unrestricted HF (bottom line)
and from HF+correlations (top line). At even values of $N$ we also show
the spin of the ground state. \\
{\bf b}: The excitation gap with spin change against $N$ (correlations are taken into account). At each N we also show the spin of the ground state and the spin of the excited state.
}
\label{fig:el1.1}
\end{figure}
\begin{figure}
\begin{center}
\includegraphics[width=.4\textwidth]{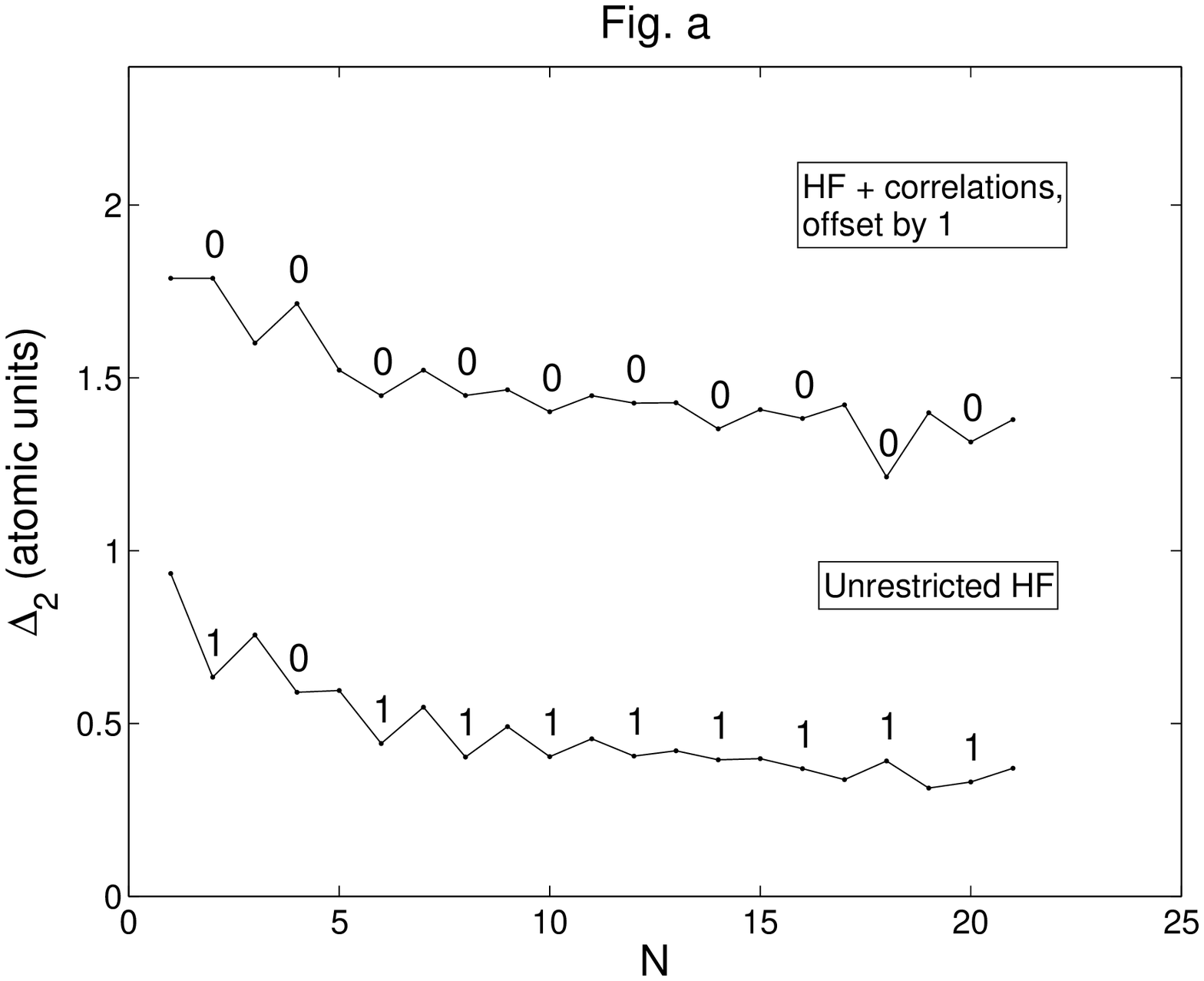}
\includegraphics[width=.4\textwidth]{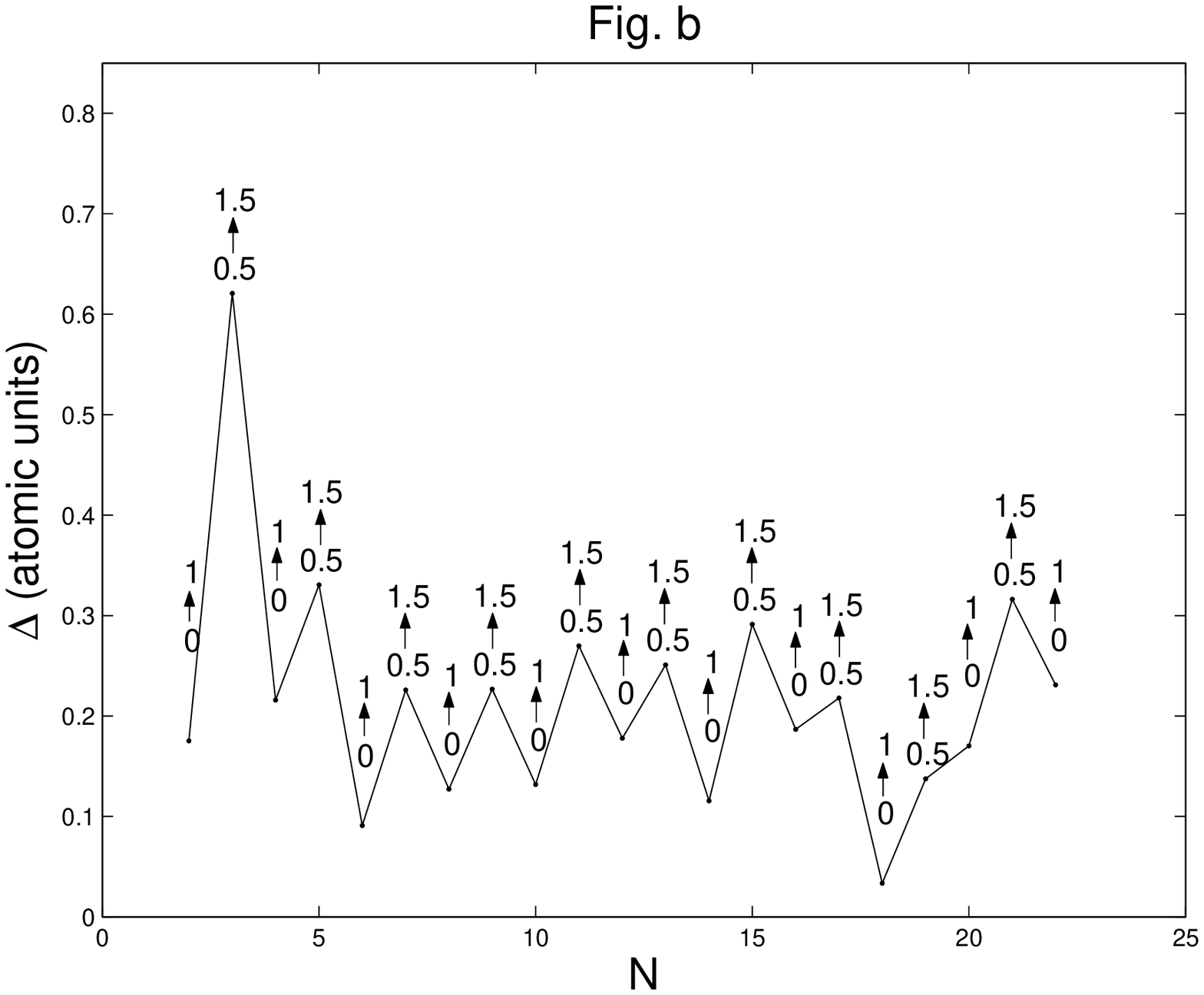}
\end{center}
\caption{The dot with ellipticity $\omega_x/\omega_y = 2$ and with
$\omega_{\textrm{at}}=\sqrt{\omega_x\omega_y}= 0.73$ which corresponds to $r_s
= 1.48 \ldots 0.99$.\\
{\bf a}: $\Delta_2$ against $N$ from unrestricted HF (bottom line)
and from HF+correlations (top line). At even values of $N$ we also show
the spin of the ground state. \\
{\bf b}: The excitation gap with spin change against $N$ (correlations are taken into account). At each N we also show the spin of the ground state and the spin of the excited state.
}
\label{fig:el2}
\end{figure}
Studies of elliptical dots are interesting from a theoretical point of
view.  Also, an elliptical deformation is a first way to model
deviations from circular geometry, which always exist in real dots. A
lateral dot, for instance, has contacts at the sides through which the
electrons tunnel onto the dot, and these alter the shape of the
confining potential.  Previous numerical studies have found that a
deformation of the dot destroys the shell
structure.\cite{austing99,lee98,HW} In the present work we investigate
a simple deformation of the dot by taking $\omega_x > \omega_y$.

Fig. \ref{fig:el1.1}a shows calculated values of $\Delta_2$ against
$N$ for a dot with the small ellipticity $\omega_x/\omega_y = 1.1$ and
with $\omega_{\textrm{at}}=\sqrt{\omega_x\omega_y}= 0.73$, which
corresponds to $r_s = 1.48 \ldots 0.99$ over the range of $N$.  The
value of the ground state spin at even values of $N$ is marked above
the lines.  The bottom line in Fig. \ref{fig:el1.1}a shows the results
of UHF calculations without correlations, and the top line (offset) in
Fig. \ref{fig:el1.1}a presents values of $\Delta_2$ obtained by taking
correlations into account.  Fig. \ref{fig:el1.1}b presents the
excitation gap with correlations included.  As in the round dot,
including correlation effects reduces the ground state spin in several
cases and hence leads to the violation of Hund's rule. In this case
the excitation gaps for these $N$ can be larger, yet the ground state
spin still changes to the minimal value.  However, most of the
$\Delta_2$ peaks given by the shell structure are still visible. The
UHF results of Fig. \ref{fig:el1.1}b are qualitatively similar,
although not identical, to the mean-field (density functional theory)
results of Austing \emph{et al.}\cite{austing99} for the
$\omega_x/\omega_y = 1.1$ dot.

Fig. \ref{fig:el2} shows the same quantities $\Delta_2$ and $\Delta$
for for dot with large ellipticity $\omega_x/\omega_y = 2$ and with
$\omega_{\textrm{at}}=\sqrt{\omega_x\omega_y}= 0.73$.  The range for
$r_s$ over $N$ remains the same, $r_s = 1.48 \ldots 0.99$.  As in
Fig. \ref{fig:w0.26}a, the spin identification within UHF is ambiguous
and should be treated with caution.  The ambiguity disappears when
correlations are included.  It is known that there are shells in a
parabolic potential with $\omega_x/\omega_y = 2$ (see
Ref. \cite{BM}). However the self-consistent potential is not
parabolic and we do not observe a shell structure in the dot, with or
without correlations.  This is evident from the gap values plotted in
Fig. \ref{fig:el2}b.  However, there is some peculiarity at
$N=17$,$18$, and $19$ and this may perhaps be considered as
reminiscent of the shell structure. The $\omega_x/\omega_y = 2$ case
was also included in the mean-field study of Austing \emph{et
al.}\cite{austing99} and no clear signature of the shell structure was
observed.

\section{Conclusions}
\label{sec:conclusions}

We have performed calculations for small quantum dots ($N \le 22$)
taking into account electron-electron correlations in the second order
of perturbation theory.  We believe that this approach is well justified
for $r_s \le 3$ when the correlation correction does not exceed a few
percent of the total energy.  To check the method we have performed a
comparison with accurate solutions for two and six electrons, in which
the method was valid to beyond $r_s=3$.  We demonstrate that
correlations are especially important for the spin structure of the
dot.  In some cases even at $r_s=1$ correlations change the spin of
the ground state from that found in Hartree-Fock and
spin-density-functional calculations. In some situations correlations
destroy Hund's rule for open electronic shells.  Finally, we observe
that correlations destroy the static spin-density waves observed in
Hartree-Fock and spin-density-functional calculations and hence
restore the spin rotational invariance of states with total spin zero.

We thank P. G. Silvestrov and A. Micolich for helpful discussions.  
We are also grateful for the computing resources provided by the Australian
Partnership for Advanced Computing (APAC) National Facility.


\begin{thebibliography}{1}

\bibitem{tarucha96} S. Tarucha, D. G. Austing, T. Honda, R. J. van der
Hage, and L. P. Kouwenhoven, Phys. Rev. Lett. \textbf{77}, 3613 (1996).
\bibitem{kouwenhoven97} L. P. Kouwenhoven, T. H. Oosterkamp, M. W. S. Danoesastro, M. Eto, D. G. Austing, T. Honda, and S. Tarucha, Science \textbf{278}, 1788 (1997).
\bibitem{alhassid00} Y. Alhassid, Rev. Mod. Phys. \textbf{72}, 895 (2000).
\bibitem{Reimann} S. M. Reimann and M. Manninen, Rev. Mod. Phys. \textbf{74}, 1283 (2002).
\bibitem{Reimann00} S. M. Reimann, M. Koskinen, and M. Manninen, Phys. Rev. B \textbf{62}, 8108 (2000).
\bibitem{KMR} M. Koskinen, M. Manninen, and S. M. Reimann, Phys. Rev. Lett. \textbf{79}, 1389 (1997)
\bibitem{austing99} D. G. Austing, S. Sasaki, S. Tarucha, S. M. Reimann, M. Koskinen, and M. Manninen, Phys. Rev. B \textbf{60}, 11514 (1999)
\bibitem{lee98} In-Ho Lee, Vivek Rao, Richard M. Martin, and Jean-Pierre Leburton, Phys. Rev. B.
\textbf{57}, 9035 (1998). 
\bibitem{RKHLM} S. M. Reimann, M. Koskinen, J. Helgesson, P. E. Lindelof, and M. Manninen, Phys. Rev. B \textbf{58}, 8111 (1998).
\bibitem{HW} K. Hirose and N. S. Wingreen, Phys. Rev. B \textbf{59}, 4604 (1999).
\bibitem{jiang01} T. F. Jiang, Xiao-Min Tong, and Shih-I Chu, Phys. Rev. B. \textbf{63}, 045317 (2001).
\bibitem{YBB} I. I. Yakimenko, A. M. Bychkov, and K. -F. Berggren, Phys. Rev. B \textbf{63}, 165309.
 (2001).
\bibitem{YL} C. Yannouleas and U. Landman, Phys. Rev. Lett. \textbf{82}, 5325 (1999).
\bibitem{YL2} C. Yannouleas and U. Landman, J. Phys.: Condens. Matter \textbf{14}, L591 (2002)\bibitem{Ov} A. W. Overhauser, Phys. Rev. \textbf{128}, 1437 (1962).
\bibitem{BM} A. Bohr and B. Mottelson, Nuclear Structure vol. 2.
\bibitem{Sneddon} I. A. Sneddon, {\it Mixed Boundary Value Problems in
Potential Theory} (Wiley, New York, 1966).
\bibitem{KS} A. A. Koulakov and B. I. Shklovskii, Phys. Rev. B \textbf{57}, 2352 (1998).
\bibitem{cances00} E. Canc\`es and C. Le Bris, Int. J. Quantum
Chem. \textbf{79}, 82-90 (2000).
\bibitem{GB} M. Gell-Mann and K. A. Brueckner, Phys. Rev. {\bf 106}, 364 (1957). 
\bibitem{com1} Strictly speaking at $N=2$, $r_s$ is not a well
defined quantity since Eq. (\ref{r1}) was derived at $N \gg
1$. Nevertheless it is convenient to use $r_s$ defined according to
(\ref{r1}) as an appropriate scaling variable.
\bibitem{Dzuba89} V. A. Dzuba, V. V. Flambaum, and O. P. Sushkov, Phys. Lett. A \textbf{141}, 147-153 (1989).
\bibitem{Blundell90} S. A. Blundell, W. R. Johnson, and J. Sapirstein, Phys. Rev. Lett. \textbf{65}, 1411 (1990).
\bibitem{Johnson04} W. R. Johnson, H. C. Ho, C. E. Tanner, and A. Derevianko, Phys. Rev. A \textbf{70}, 014501 (2004).


\end{thebibliography}
\end{document}